\documentclass{article}

\usepackage{amsfonts,amssymb,amsmath, amsthm, color, a4wide}
\usepackage{mathtools}
\usepackage{mathrsfs}
\usepackage{hyperref}
\usepackage{graphicx}
\usepackage{epstopdf}
\usepackage{tikz}
\usetikzlibrary{shapes.geometric, arrows}
\usetikzlibrary{positioning}

\tikzstyle{inout} = [ellipse, minimum width=3cm, minimum height=1cm, text centered, draw=black]
\tikzstyle{layer} = [rectangle, rounded corners, minimum width=3cm, minimum height=1cm,text centered, draw=black]
\tikzstyle{maxpool} = [rectangle, minimum width=2cm, minimum height=1cm, text centered, draw=black]
\tikzstyle{arrow} = [thick,->,>=stealth]

\usepackage[round]{natbib}
\setcitestyle{notesep={: }}
\setcitestyle{aysep={,}}

\numberwithin{equation}{section}

\usepackage{authblk}

\renewenvironment{abstract}{%
	\vspace{6pt}%
	\begin{center}%
		\begin{minipage}{320pt}%
			\small%
			\begin{center}%
				\textbf{Abstract}%
			\end{center}%
		}{\end{minipage}\end{center}}
\newcommand{\keywords}[1]{%
	\begin{center}%
		\begin{minipage}{320pt}%
			\textit{Keywords:}~{#1}
		\end{minipage}%
	\end{center}%
}

\title{Using neural networks to estimate parameters in spatial point process models}
\author[1]{Ninna Vihrs}
\affil[1]{Department of Mathematical Sciences, Aalborg University}

\begin{document}

\maketitle

\begin{abstract}
In this paper, I show how neural networks can be used to simultaneously estimate all unknown parameters in a spatial point process model from an observed point pattern. The method can be applied to any point process model which it is possible to simulate from. Through a simulation study, I conclude that the method recovers parameters well and in some situations provide better estimates than the most commonly used methods. I also illustrate how the method can be used on a real data example.
\end{abstract}

\section{Introduction}
Briefly, a point process $X$ may be defined as a countable random subset of $\mathbb{R}^d$. Usually, a realisation of the process is only observed within a bounded set $W\subset \mathbb{R}^d$. A common problem is to fit a parametric spatial point process model to a realization $x$. This can be a difficult problem since the likelihood function is intractable except in the very simple case of a Poisson process. Many alternative approaches have thus been suggested including estimation based on pseudo-likelihood, composite likelihood, and minimum contrasts \citep[see the review in][]{JMRW2017}. However, it is possible to define meaningful spatial point process models for which both the intensity and other moment characteristics of $X$, the density, and the Papangelou conditional intensity \citep[see e.g.][]{textbook} are not expressible in closed form. Then, the above methods are not feasible. An example of such a point process model is the LGCP-Strauss process presented in \citet{LGCPStrauss} where the authors found it necessary to consider parameter estimation in a Bayesian setup because it was then possible to use the method of approximate Bayesian computations (ABC) which is based entirely on the ability to simulate under the model (see e.g.\ the overview of some ABC methods in \citet{ABCoverview}).

The purpose of this paper is to explore the possibility to estimate parameters in spatial point process models by using neural networks. The idea is to consider the estimation problem as a prediction problem where parameters of the model are to be predicted from a realization. This prediction problem can be handled with machine learning methods, such as neural networks, trained on a suitable training data set. Thus the only requirement for this approach is to be able to construct a number of training cases consisting of values for the unknown parameters and realizations of the model corresponding to these parameter values. If it is known how to simulate from the model, the training data set for the chosen machine learning method can be constructed from simulations of the model. Thus the approach, like ABC, only requires the model to be equipped with a feasible simulation procedure. The idea is somewhat similar to the concept behind the ABC technique in \citet{abcrf} where random forests are used to predict mean, variance, and quantiles in the posterior distribution. However, I instead use neural networks for the prediction task, and is furthermore only interested in obtaining point estimates for the unknown parameters and do not attempt to get knowledge about a posterior distribution. \citet{physical_modelling} have used neural networks to predict parameters in acoustic physical modelling, but the idea has to the best of my knowledge not been explored in relation to spatial point process models. I explain the suggested approach for using neural networks to obtain point estimates of unknown parameters of spatial point process models in Section~\ref{sec:NNet}.

Neural networks have proven useful in many different prediction problems and are well suited to handle many different types of input including images and curves. Their ability to handle different input formats is an advantage when attempting to pass information about a relatively complex data structure like a point pattern. In Section~\ref{sec:sumstats}, I discuss how to pass information about a point pattern to a neural network aiming at predicting unknown parameters; I decide on summarising important aspects of the point pattern by means of a functional summary statistic and then passing this information to the neural network, thereby using the possibility to handle input data in the form of curves.

As I mention above, there are many different estimation procedures for spatial point process models, and which one it is preferable to use depends on the type of model and the theoretical knowledge available for that class of models. A clear advantage of simulation based methods, like the neural network approach in this paper, is that they are generally applicable to all point process models for which it is possible to generate realizations. Since the ability to simulate from the model must be considered necessary for any model of practical value, this requirement is not very restrictive. Good simulation based methods thus allow us to use almost any type of spatial point process model without being limited by lack of theoretical knowledge when it comes to parameter estimation. Another clear advantage of the suggested neural network approach in this paper is that all unknown parameters can always be estimated simultaneously, which is not always the case in traditional estimation procedures. For instance, I consider the example of a Strauss process in Section~\ref{sec:strauss} where parameters are usually fitted with the method of maximum pseudo likelihood estimation, but the Strauss process contains an interaction radius $R$, and this parameter cannot be estimated alongside the other parameters when using maximum pseudo likelihood estimation. Finally, I show through the simulation study in Section~\ref{sec:simstudy} that the suggested neural network approach recovers parameters well, and compared to the most commonly used estimation procedures it gives either better or similar results. 

All statistical computations in this paper were made with the open source software \texttt{R} version 4.0.2 \citep{R}. The \texttt{R}-packages \texttt{ggplot2} version 3.3.2 \citep{ggplot}, \texttt{spatstat} version 2.1-0 \citep{spatstat}, and \texttt{keras} version 2.3.0.0 \citep{keras, keras_book} were used to make plots, handle spatial point patterns, and train neural networks, respectively. The $R$-scripts I wrote for the simulation studies and data example in this paper are available in the ancillary files.

\section{The neural network approach}
\label{sec:NNet}

In this section, I explain the suggested neural network approach to parameter estimation in spatial point process models. I restrict attention to models without covariates and leave the case of inhomogeneous models to future research. The objective is to train a neural network to predict the values of parameters in a chosen point process model based on a realisation from the process. When the neural network has been trained, it can be used to estimate the parameters of the point process model based on an observed point pattern $x_{\mathrm{obs}}$. 

\subsection{Considerations regarding training data}\label{sec:sumstats}
In order to train the neural network, training data is constructed by simulating a number of point patterns from the chosen model for different values of the parameters. Neural networks are known to be able to take input data in many forms including pixel images and sequences. One way to pass a point pattern dataset to a neural network would be to represent it as a pixel image where the pixel values corresponds to the number of points within the pixel. I tried to send data in this form through a 2-dimensional convolutional neural network, which is a good choice for handling image input, but this method seemed to be less successful than summarising the point pattern dataset with functional summary statistics as explained below. This may be because the behaviour of a point pattern at a very small scale is important for estimating some parameters accurately, and such information was lost in the discretization of the pattern but not when summarising aspects of the pattern with a functional summary statistic. The need to choose an appropriate summary statistic brings some arbitrariness and subjectivity to the method which would not have been the case if using the entire point pattern as input, but it is not uncommon to estimate parameters in spatial point process models based on some functional summary statistic as this is also done in the popular method of minimum contrast estimation (see Section~\ref{sec:LGCP}). In this paper, I therefore choose to use some appropriate summaries calculated from the point pattern as input to the neural network.

A common way to summarise many important aspects of a point pattern is by means of functional summary statistics where I briefly describe some common choices here and refer the reader to \citet{spatstat} for more details. A common choice is Ripley's $K$-function which depends on an inter point distance $r$. Assuming stationarity of the point process, if $\rho$ is the intensity of the process, the interpretation of $K$ is that $\rho K(r)$ is the expected number of further points falling in a ball with radius $r$ centered at a typical point of the process. One often considers its transformation $L(r)=\sqrt[d]{K(r)/\omega_{d}}$ where $\omega_d$ is the volume of a $d$-dimensional unit sphere. It is known that $L(r)=r$ in case of a stationary Poisson process, which is the case of complete spatial randomness. If $L(r)-r<0$ ($L(r)-r>0$), it is usually interpreted as the point process exhibiting regularity/repulsion (clustering/attraction) at interpoint distances $r$. Non-parametric estimates of $K$ and $L$ from a point pattern $x=\{x_1,\ldots,x_n\}$ on an observation window $W$ are
\begin{align*}
\widehat{K}_{x}(r)=\frac{|W|}{n(n-1)}\sum_{i=1}^{n}\sum_{j\neq i, j=1}^{n}1[\|x_i-x_j\|\leq r]e_{ij}(r), \quad \widehat{L}_{x}(r)= \sqrt[d]{\widehat{K}_{x}(r)/\omega_{d}}
\end{align*}
where $|W|$ is the Lebesgue measure of $W$ and $e_{ij}(r)$ is an edge correction weight to account for the unobserved points outside $W$. \citet{spatstat} noted that it is not so important which edge correction method to use as long as some correction is used; I used Ripley's isotropic correction \citep{Ripley88, iso2}. I use $\hat{L}(r)-r$ as input to the neural network since parameters of point processes are usually related to regularity and clustering, and it was found to give better results than using Ripley's $K$-function or the $L$-function directly, which may suggest that this transformation of $K$ allows the neural network to learn more efficiently. Note however that the suggested neural network approach can easily be used with a different functional summary statistic as input.

Other popular summary functions for point processes include the so-called $F$-, $G$-, and $J$-functions defined for a stationary point process $X$ by
\begin{align}
F(r)&=P(X\cap b(0,r)\neq\emptyset),\\
G(r)&=P((X\setminus \{u\})\cap b(u,r)\neq \emptyset\mid u\in X),\quad \text{and}\\
J(r)&=\frac{1-G(r)}{1-F(r)}, \quad F(r)<1,\label{eq:J}
\end{align}
where $b(u,r)$ is the ball centered at $u$ with radius $r$, see e.g.\ \citet{textbook} and the references therein. Stationarity implies that the definition of $G(r)$ does not depend on the choice of $u$. The $F$-, $G$-, and $J$-functions are however not considered further as input to the neural network, since they can usually only be estimated reliably for a smaller range of $r$-values than $K$, and this was found to be a disadvantage for the example in Section~\ref{sec:LGCPStrauss} where large scale properties had to be summarised.

The number of points in the point pattern was also included in the input to the neural network, since this is important knowledge regarding some parameters of most point process models which the $L$-function is generally not able to provide. 

\subsection{The suggested neural network approach}\label{sec:NN_method}
The suggested method for estimation is as follows:
\begin{enumerate}
\item Choose a homogeneous spatial point process model $M(\theta)$ with unknown parameters $\theta=(\theta_1,\ldots,\theta_k)$, a number of training cases $n_{\mathrm{train}}$, and optionally a number of test cases $n_{\mathrm{test}}$.
\item Construct training data:
\begin{enumerate}
\item \label{train1} For $i=1,\ldots,n_{\mathrm{train}}$, sample the parameters $\tilde{\theta}^{i}=(\tilde{\theta}_1^{i},\ldots,\tilde{\theta}_k^{i})$ from some pre-chosen distribution for $\theta$. In this paper, I sample the $\tilde{\theta}_j^{i}$'s independently and uniformly on bounded intervals. 
\item For $i=1,\ldots,n_{\mathrm{train}}$, sample $\tilde{x}^{i}$ from $M(\tilde{\theta}^{i})$.
\item \label{train3} Choose some values $r_1,\ldots,r_m$. For $i=1,\ldots,n_{\mathrm{train}}$ calculate the functional summary statistic $L_i=(\widehat{L}_{\tilde{x}^i}(r_1) - r_1,\ldots,\widehat{L}_{\tilde{x}^i}(r_m) - r_m)$ and $n(\tilde{x}^i)$ where $n(\cdot)$ is the number of points in a point pattern.
\item \label{standardize} Standardize each component of $\{L_i,n(\tilde{x}^i),\tilde{\theta}_1^{i},\ldots,\tilde{\theta}_{k}^{i}\}_{i=1}^{n_{\mathrm{train}}}$, by subtracting the mean and dividing by the standard deviation (for $\{L_i\}_{i=1}^{n_{\mathrm{train}}}$ the mean and standard deviation were calculated both over all $n_{\mathrm{train}}$ simulations and over all $m$ values for $r$ meaning that all values of $\{L_i\}_{i=1}^{n_{\mathrm{train}}}$ were scaled by the same amount.) After standardization the training data is $\{L_i,n(\tilde{x}^i),\tilde{\theta}_1,\ldots,\tilde{\theta}_{k}\}_{i=1}^{n_{\mathrm{train}}}$.
\end{enumerate}
\item (Optional) Construct test data:
\begin{enumerate}
\item Construct $n_{\mathrm{test}}$ test cases $\{L_i,n(\tilde{x}^i),\tilde{\theta}_1^{i},\ldots,\tilde{\theta}_{k}^{i}\}_{i=1}^{n_{\mathrm{test}}}$ with the same procedure as in items \ref{train1}--\ref{train3}.
\item Scale the test data according to item \ref{standardize}, i.e. subtract the means and divide by the standard deviations calculated in item \ref{standardize}.
\end{enumerate}
\item \label{fit_nnet} Use the training data to train a neural network to predict $\theta$.
\item (Optional) Send the test data through the trained neural network, and asses its predictive performance.
\item Calculate $L_\mathrm{obs}=(\widehat{L}_{x_{\mathrm{obs}}}(r_1) - r_1,\ldots,\widehat{L}_{x_{\mathrm{obs}}}(r_m) - r_m)$ and $n(x_{\mathrm{obs}})$; scale these according to item \ref{standardize}; feed them to the trained neural network; and return the (rescaled) prediction $\hat{\theta}$ as the estimated vector of parameters.
\end{enumerate}
Regarding the choice of the values $r_1,\ldots,r_m$ in item \ref{train3} there is a sensible default in the \texttt{spatstat} implementation for estimating $L(r)$, which I used. Even though it is optional to construct test data, I strongly recommend to do this in order to asses the performance of the method in a given situation.

As I write in item \ref{train1}, I sample each parameter in the training data uniformly on a bounded interval, in which case there should be strong reasons to believe that the parameters corresponding to the observed point pattern fall within these intervals. Otherwise, the trained neural network cannot be expected to do well for the observed point pattern. Note however, that it is possible to consider wide intervals of the parameters, so it is not necessary to have very specific knowledge about the ranges of the parameters. It will usually be possible to obtain some range for each parameter by combining knowledge about the effect of the parameters in the model with a preliminary investigation of the point pattern, which may include interpreting some functional summary statistics and looking at some simulations. I give an idea of how this could be done for the example in Section~\ref{sec:dataex}. 

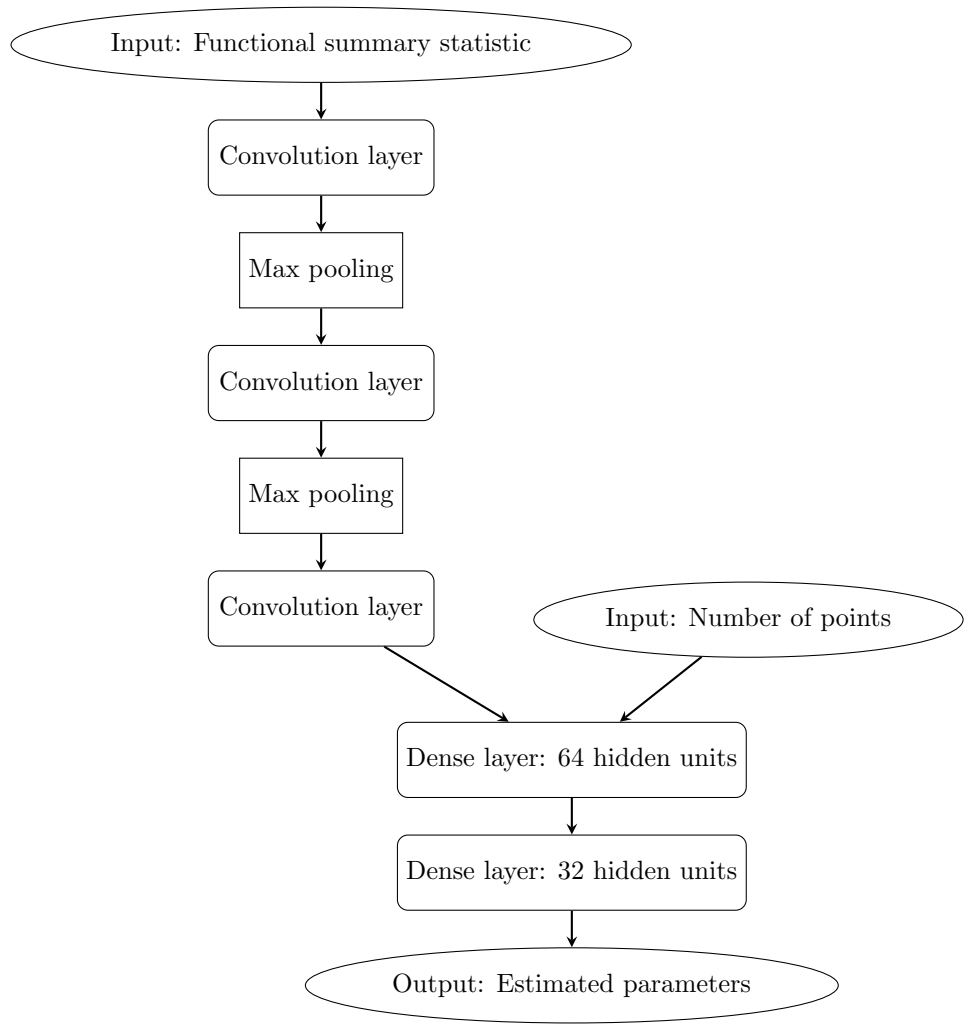
\begin{figure}[htp!]
\centering
\begin{tikzpicture}[node distance=1.5cm]
\node (input) [inout] {Input: Functional summary statistic};
\node (conv1)[layer, below of = input]{Convolution layer};
\node (max1)[maxpool, below of = conv1]{Max pooling};
\node (conv2)[layer, below of = max1]{Convolution layer};
\node (max2)[maxpool, below of = conv2]{Max pooling};
\node (conv3)[layer, below of = max2]{Convolution layer};
\node (dense1)[layer, below right = 1cm and -0.5cm of conv3]{Dense layer: 64 hidden units};
\node (inputN)[inout, above right = 1cm and -2cm of dense1]{Input: Number of points};
\node (dense2)[layer, below of = dense1]{Dense layer: 32 hidden units};
\node (output)[inout, below of = dense2]{Output: Estimated parameters};
\draw [arrow] (input) -- (conv1);
\draw [arrow] (conv1) -- (max1);
\draw [arrow] (max1) -- (conv2);
\draw [arrow] (conv2) -- (max2);
\draw [arrow] (max2) -- (conv3);
\draw [arrow] (conv3) -- (dense1);
\draw [arrow] (inputN) -- (dense1);
\draw [arrow] (dense1) -- (dense2);
\draw [arrow] (dense2) -- (output);
\end{tikzpicture}
\caption{Visual overview of the neural network architecture.}
\label{fig:NNet}
\end{figure}

The neural network architecture which I chose to use in item \ref{fit_nnet} is illustrated in Figure~\ref{fig:NNet}.
The functional summary statistic $\widehat{L}(r)-r$, which constitutes a sequence, is send through a number of 1-dimensional convolution layers and max pooling operations, which is a good way to handle sequenced data.

A 1-dimensional convolution layer takes as input a number of sequences say $s^{i}=(s^{i}_1,\ldots, s_{k}^{i}),$ $i=1,\ldots, m$, and returns $p$ sequences of the form $\tilde{s}^{i}=(\tilde{s}^{i}_1,\ldots, \tilde{s}_{k-(q-1)}^{i})$, $i=1,\ldots,p$, where $p$ is some chosen number and $\tilde{s}^{i}_j=f(b_{i} + \sum_{l=0}^{q-1}\sum_{h=1}^{m}a^{i}_{lh}s^{h}_{j + l})$ for some activation function $f$, chosen size $q$, constant $b_i$, and weights $a^{i}_{lh}, l=0,\ldots,q-1, h = 1,\ldots,q$. I used $p=64$ in each convolution layer and $q=7$, so in the first convolution layer $m=1$ (the functional summary statistic is just one curve) and in the subsequent layers $m=64$ (the output from the previous layer constitutes 64 curves).

The max pooling operation used between the convolution layers splits every input sequence into sub-sequences of a specified length (I chose 5) and replaces each sub-sequence with its highest value yielding a new sequence usually of much smaller size.

After the convolution layers, the output is fed to two densely connected layers (the output is now considered as individual values instead of sequences) together with the number of points in the observed point pattern. In a densely connected layer which gets input values $I_1,\ldots, I_n$, the output is $O_1,\ldots, O_h$ where $h$ is some chosen number of hidden units and $O_i=f(b_i + \sum_{j=1}^{n}a^{i}_{j}I_j)$ for some activation function $f$, constant $b_i$, and weights $a^{i}_j, j=1,\ldots,n$. The final output of the network is a prediction of the unknown parameters of the spatial point process model based on the functional summary statistic and number of points which was given as input to the model. This final layer is actually also a dense layer where the activation function is just the identity and the number of output values corresponds to the number of unknown parameters in the spatial point process model which is to be estimated.

As the activation function I used $f(\cdot)=\mathrm{relu}(\cdot)=\max(0, \cdot)$ both for the convolution and dense layers, which is a very common choice for the activation function in neural networks. All the above mentioned weights of the linear combinations taken in the neural network and the constants $b_i$ constitute the unknown parameters of the neural network, which should be learned based on the training data. For some details about how these unknown parameters of the neural network were learned see Appendix~\ref{app:NN_learn}. For more information about neural networks and how to use them in \texttt{R} see e.g.\ \citet{keras_book}.

I also tried to use a network only with densely connected layers, which is much faster to train, but it generally gave poorer results than including the convolution layers.

\section{Simulation study for examples of point process models}
\label{sec:simstudy}
In this section, I consider three classes of parametric spatial point process models as examples: log-Gaussian Cox processes (LGCP) \citep{LGCP}, Strauss processes \citep{Strauss, noteStrauss}, and LGCP-Strauss processes \citep{LGCPStrauss}. I briefly define these in the following subsections and refer to the above references for more details about these models. The preferred method for estimating parameters in spatial point process models depends on the type of model. Through simulations, I assess the accuracy of estimates obtained with the neural network approach and compare this to the most commonly used estimating procedure in each case, which I briefly describe in each of the following subsections. I will not go into details about simulation methods and instead refer to \citet{textbook} and \citet{spatstat}. In this section, $W$ is always a 2-dimensional unit square. Considerations about how many simulations to use for the training data in each example are provided in Appendix~\ref{app:trainingdata}, which also shows histograms of the number of points in the simulations in the training data sets.

\subsection{LGCP processes}\label{sec:LGCP}
An LGCP is a popular process for modelling aggregation in spatial point patterns. It is driven by a stochastic intensity $Z = \exp(Y)$ where $Y$ is a Gaussian random field with mean function $m$ and covariance function $c(u,v)$. I use $m=\mu$ for a constant $\mu$ and an exponential covariance function $c(u,v)=\sigma^2\exp(-\|u-v\|/s)$ with unknown parameters $\sigma^2$ and $s$. The model is then stationary, and the vector of unknown parameters which are to be estimated is $(\mu, \sigma^2, s)$. Note that if $\sigma^2=0$, $s$ becomes irrelevant, so any estimation procedure may be expected to struggle with estimating $s$ if $\sigma^2$ is small. 

Seemingly, the most common way to estimate parameters in Cox processes is to use the method of minimum contrast estimation \citep{mincon83, mincon84}, although composite likelihood estimation \citep{complik} is a popular alternative. I have chosen to compare my method to that of minimum contrast estimation. In minimum contrast estimation, the theoretical value of a summary function (e.g.\ Ripley's $K$-function) is compared to a non-parametric estimate of it. I used this method to estimate $\sigma^2$ and $s$ based on an observation $x$ and Ripley's $K$-function, which depends on $\sigma^2$ and $s$, by finding the values of $\sigma^2$ and $s$ which minimize
\begin{equation*}
\int_{a_1}^{a_2}|K(r)^{q}-\widehat{K}_{x}(r)^{q}|^{p}\,\text{d}r
\end{equation*}
for some user specified $0\leq a_1<a_2$ and exponents $p$ and $q$. Subsequently, $\mu$ can be estimated from the unbiased estimation equation of the intensity $\hat{\rho}=n(x)/|W|$ by using that $\rho = \exp(\mu + \sigma^2/2)$ for my considered model. For finding the minimum contrast estimates I used the function \texttt{kppm} from \texttt{spatstat} with the default settings, which include $p = 2$ and $q = 1/4$.

\begin{figure}[ht!]
\centering
\includegraphics[scale=1]{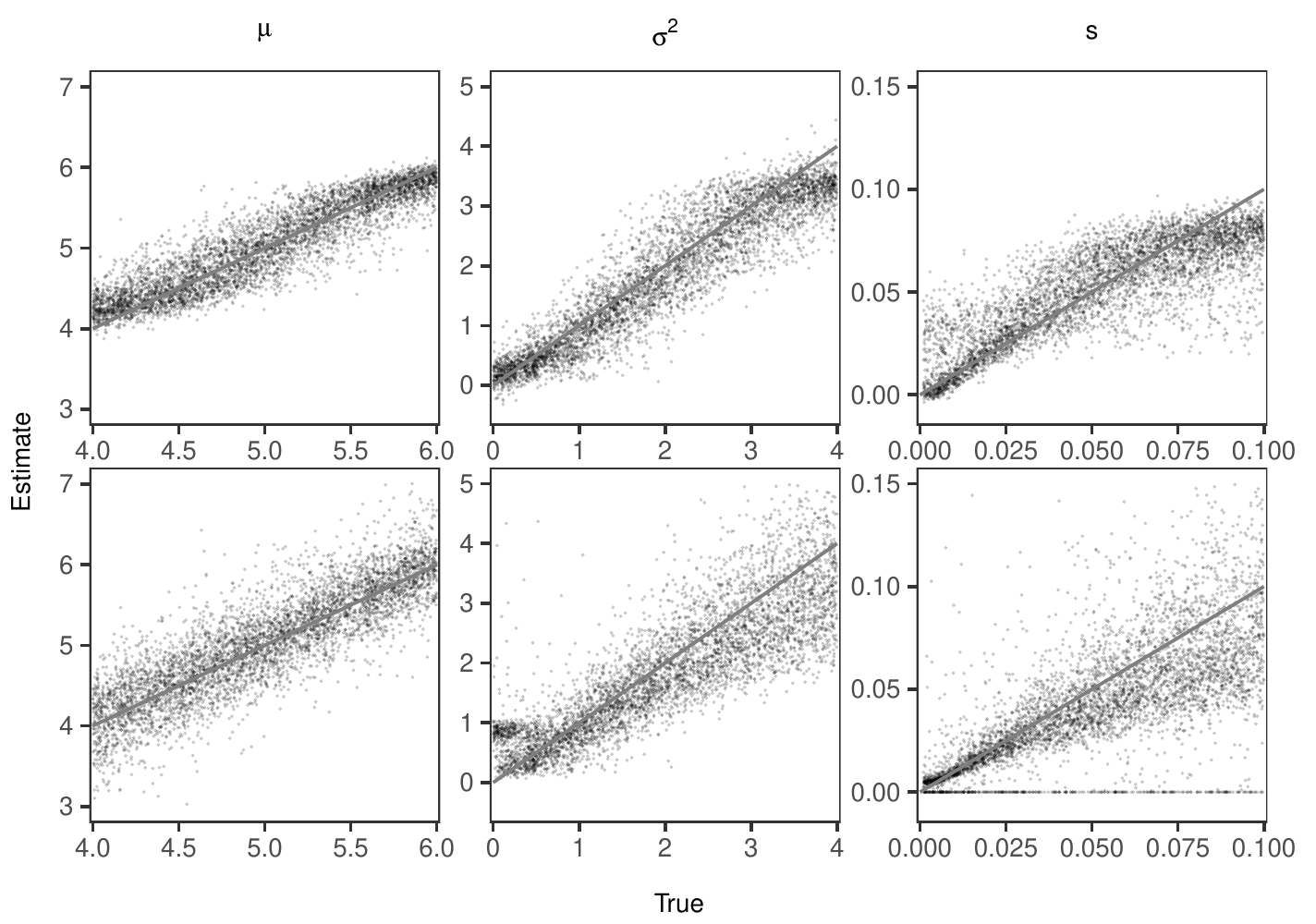}
\caption{Estimated parameters obtained with the neural network approach (top row) and minimum contrast estimation (bottom row) plotted against true parameters in an LGCP. The solid gray line is the identity line. In the case of minimum contrast estimation, 15, 43, and 77 cases where the estimate of $\mu$ was below 3, $\sigma^2$ was above 5, and $s$ was above 0.15, respectively, were omitted from the respective plots; the smallest estimate of $\mu$ was $-0.174$, the highest estimate of $\sigma^2$ was $10.7$, and the highest estimate of $s$ was $78{,}738.35$.}
\label{fig:LGCP_estimates}
\end{figure}

\begin{figure}[ht!]
\centering
\includegraphics[scale=1]{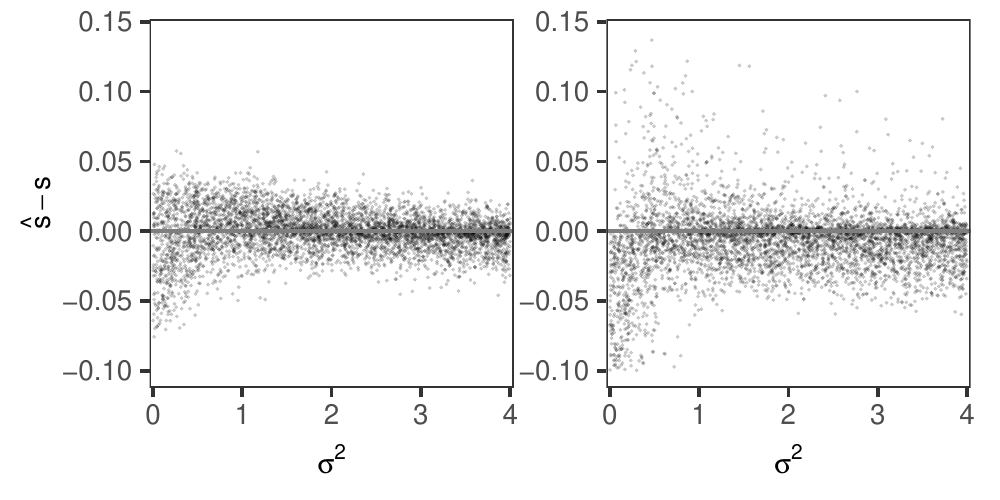}
\caption{Estimate of $s$ minus the true value plotted against $\sigma^2$ for a log-Gaussian Cox process. In the plot to the left, estimates were obtained with the neural network approach; in the plot to the right, estimates were obtained with minimum contrast estimation. In the case of minimum contrast estimation, 34 cases where the error fell outside the showed range were omitted.}
\label{fig:LGCP_special}
\end{figure}

\begin{figure}[ht!]
\centering
\includegraphics[scale=1]{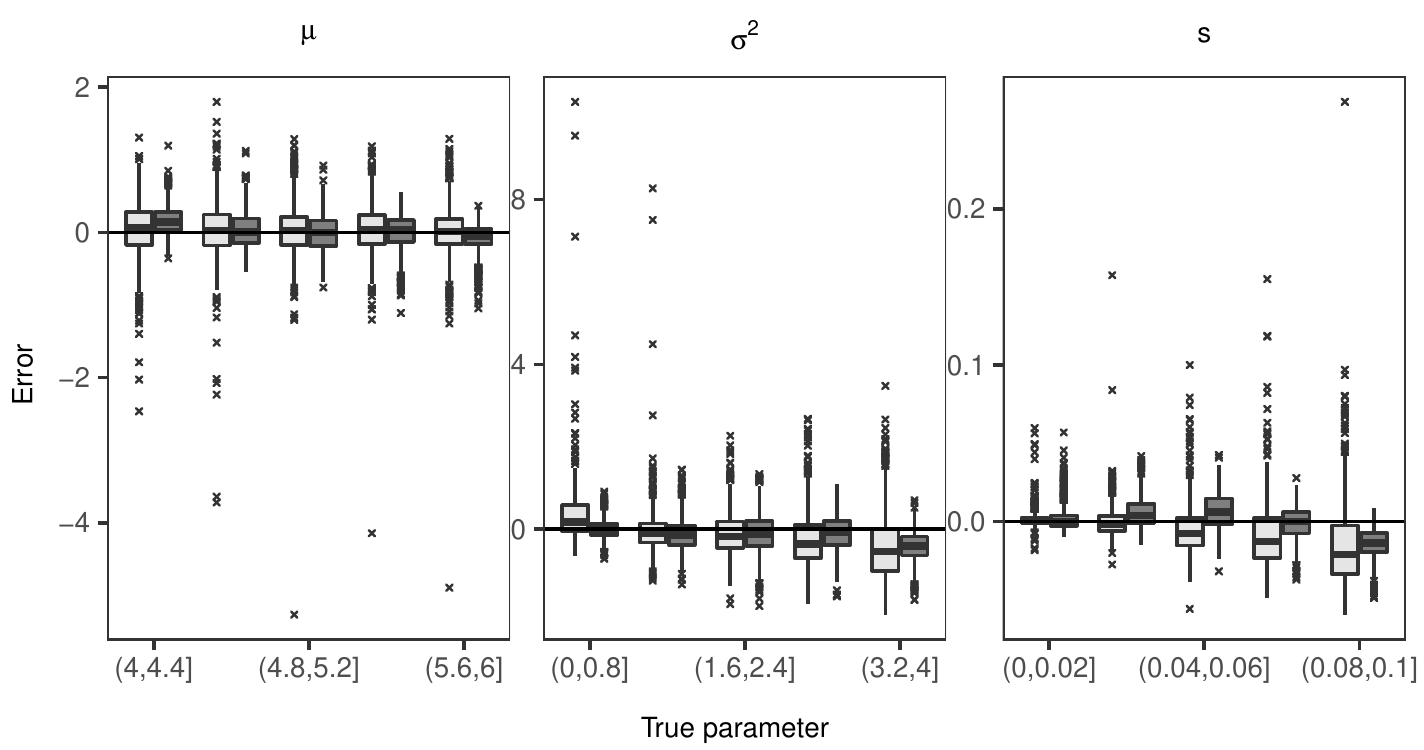}
\caption{Boxplot of the errors (estimated minus true value) for the parameter of an LGCP stated at the top of each plot. The estimates were obtained with the neural network approach (dark gray) and minimum contrast estimation (light gray). For the parameter $s$, cases where $\sigma^2>1$ were omitted.}
\label{fig:LGCP_boxplot}
\end{figure}

I made the training data for the neural network approach based on $10{,}000$ simulations of an LGCP with parameters sampled uniformly in the intervals $\mu\in (4,6)$, $\sigma^2\in (0, 4)$, and $s\in(0.001,0.1)$. For justification of the number of simulations in the training data see Appendix~\ref{app:trainingdata}. I made further $5{,}000$ simulations for a test set, and Figures~\ref{fig:LGCP_estimates}--\ref{fig:LGCP_boxplot} show some plots for the estimated parameters obtained with the neural network approach and the method of minimum contrast estimation.
In the case of minimum contrast estimation, some extreme estimates were omitted in Figures~\ref{fig:LGCP_estimates} and \ref{fig:LGCP_special}, see the captions for more details. Overall, there is less variation in the error of the estimates obtained with the neural network approach compared to the method of minimum contrast estimation, especially when the true parameter is high. Furthermore, the neural network approach does not give the same kind of extremely wrong estimates as sometimes seen with minimum contrast estimation, probably because it has only seen training data with parameters in the same intervals as in the test set.

For $\mu$, both methods recover the parameter well, but the neural network approach has unlike the method of minimum contrast estimation a slight tendency to overestimate the parameter when the true value is small.

For $\sigma^2$, the neural network approach recovers the parameter well when the true value is less than three, especially when the true value is very small where it also performs considerably better than minimum contrast estimation since the latter quite often seems to estimate the parameter to be near one when it is in fact near zero. When the true parameter is high, both methods have a tendency for underestimation. However, in the case of minimum contrast estimation, this tendency starts to be clear when the true value gets above circa $2.5$ whereas it for the neural network approach only starts to be clear when the true value gets above circa $3.5$.

For $s$, Figure~\ref{fig:LGCP_special} shows that both methods as expected struggle to recover $s$ when $\sigma^2$ is near 0. In this case, the neural network approach has a tendency to estimate $s$ to be near the mean in the training data whereas the method of minimum contrast has a tendency to estimate it to be near 0. There is no reason to prefer any of these strategies above the other, so for a more fair comparison, cases where $\sigma^2<1$ has been excluded from the plot for $s$ in Figure~\ref{fig:LGCP_boxplot} and we see that the excluded cases include the most extreme estimates of $s$ achieved with minimum contrast estimation. Both methods recover $s$ well when the true value is small (and $\sigma^2$ is not near 0). When the true value of $s$ gets above circa $0.3$, the method of minimum contrast develops a tendency for underestimation, which gets more severe as $s$ increases, and the neural network approach starts to slightly overestimate $s$ until the true value gets above circa 0.8 after which it also underestimates $s$, but not as severely as minimum contrast estimation. 

\begin{figure}[ht!]
\centering
\includegraphics[scale=1]{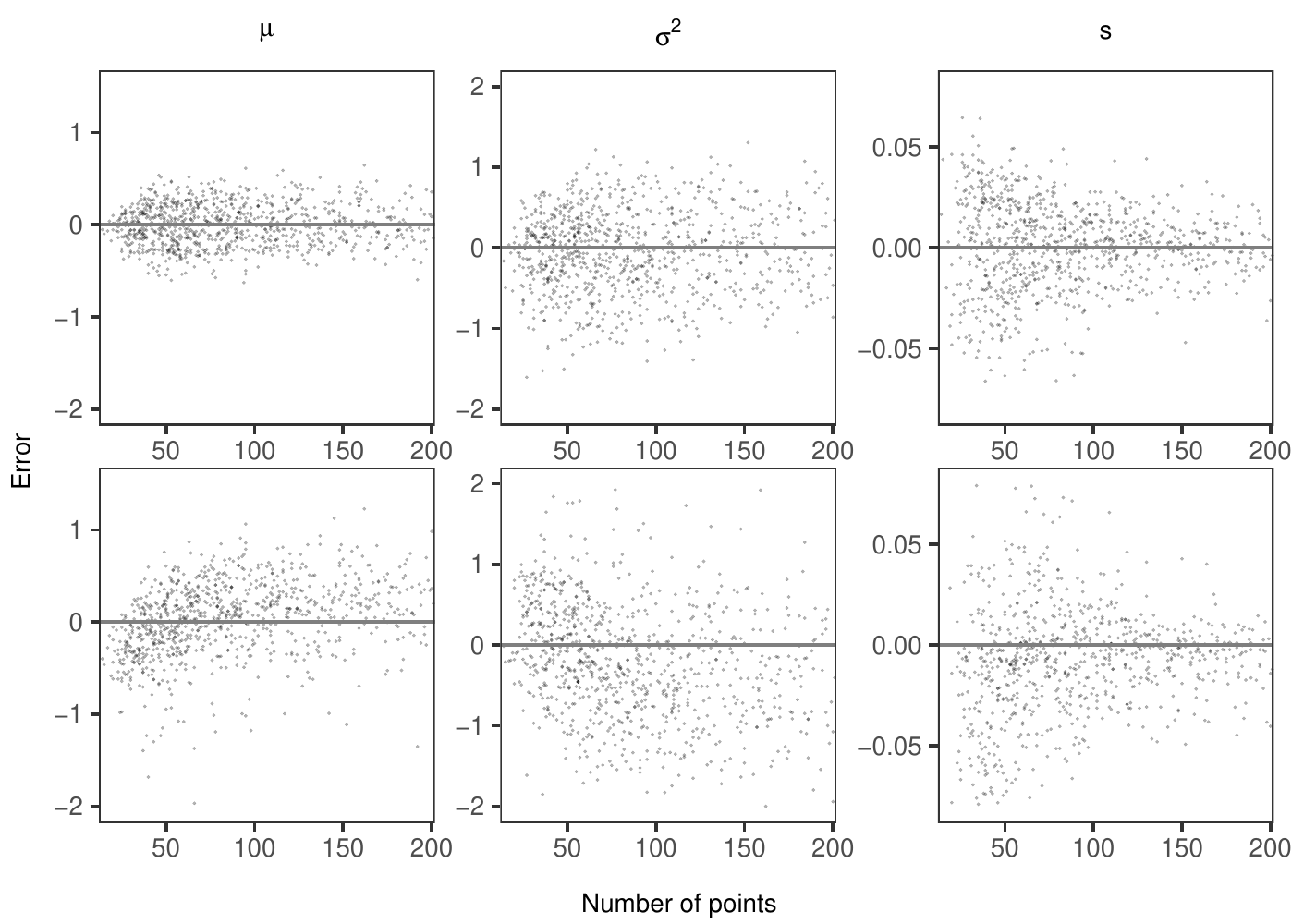}
\caption{Estimated minus true value plotted against the number of points in the point pattern. The parameter of the LGCP is stated at the top of each column. In the top row, estimates were obtained with the neural network approach; in the bottom row, estimates were obtained with minimum contrast estimation. In the case of minimum contrast estimation, 2 cases of $\mu$, 19 cases of $\sigma^2$, and 100 cases of $s$ where the error fell outside the showed ranges were omitted.}
\label{fig:error_vs_npoints}
\end{figure}

In order to asses how the method performs on point patterns with few points, I made a second simulation study where I considered $\mu\in(3,4)$ for the training and test data. I used $1{,}000$ simulations in the test set and everything else was as above. Figure~\ref{fig:error_vs_npoints} shows the errors of the estimates obtained with the neural network approach and minimum contrast estimation for each parameter for the test cases where the number of points was below 200. 
This shows that it is mainly the estimation of $s$ which benefits from more points in the point pattern. It is also seen that minimum contrast estimation has a tendency to underestimate $\mu$ and $s$ and overestimate $\sigma^2$ if there are very few points in the point pattern, but the neural network approach shows no such tendencies. 

\subsection{Strauss processes}\label{sec:strauss}
A Strauss process is a popular model for regularity. A Strauss process defined on a bounded set $S\subset \mathbb{R}^d$ has density $f(x)\propto \beta^{n(x)}\gamma^{S_R(x)}$ with respect to a unit rate Poisson process for $x=\{x_1,\ldots ,x_n\}\subset S$ where $n(x)$ is the number of points in $x$, $S_R(x)=\sum_{i<j}1[\|x_i-x_j\|\leq R]$ is the number of $R$-close pairs, and the unknown parameters are $\beta >0$, $\gamma\in[0,1]$, and $R\geq 0$. I assume that $W\subset S$ but that $S$ is unknown, so when simulating from the Strauss process, I simulate it on an extended window determined by the default settings in the function \texttt{rmh} from \texttt{spatstat}, which in this case is to add a margin of width $2R$ around all sides of the square $W$. Note that if $\gamma = 1$, $R$ becomes irrelevant; and if $R=0$, $\gamma$ becomes irrelevant. Both these special cases collapses into the same model namely a homogeneous Poisson process. 

The density of the Strauss process involves an intractable normalising constant, so instead of using maximum likelihood estimation it is more common to use maximum pseudo likelihood estimation \citep{MPLEBesag, Ripley88, MPLEJensenMoller, MPLEBaddeleyTurner}, which is known to be a fast and reliable alternative. The pseudo likelihood function for an observed point pattern $x$ is
\begin{equation*}
PL_{A}(\theta)=\exp\left (-\int_{A}\frac{f(x\cup \{u\})}{f(x)}\,\text{d}u\right )\prod_{u\in x\cap A}\frac{f(x)}{f(x\setminus\{u\})}
\end{equation*}
for some set $A\subset S$ chosen in order to account for edge effects. The pseudo likelihood function is maximised in order to find the maximum pseudo likelihood estimate of an unknown parameter vector $\theta$ of the density $f$. This method is particularly tractable if the model is on exponential family form, that is the unnormalised  density $h$ is of the form $h(x)=\exp(t(x)\theta^T)$ for a canonical parameter vector $\theta$ and canonical statistic $t(x)$. This is the case for the Strauss process if $R$ is given with $\theta=(\log(\beta),\log(\gamma))$ and $t(x)=(n(x),S_{R}(x))$. Thus, maximum pseudo-likelihood estimation can easily be used to obtain estimates of $\beta$ and $\gamma$. In order to also obtain an estimate for $R$, the method of profile maximum pseudo likelihood can be used in the following way: consider a finite set $R_1,\ldots,R_k$ of possible values for $R$, find the maximum pseudo likelihood estimates $\hat{\beta}_{i}$ and $\hat{\gamma}_i$ of $\beta$ and $\gamma$ given $R=R_{i}$, and choose the combination of parameters $(\hat{\beta}_{i},\hat{\gamma}_i, R_i)$ which gives the highest value of $PL_{A}$. For finding the profile maximum pseudo likelihood estimates I used the function \texttt{profilepl} from \texttt{spatstat} where I forced the method to yield a valid model ($\gamma\in[0,1]$) and considered $50$ equally spaced values of $R$ in the interval $[0.001, 0.05]$.

\begin{figure}[ht!]
\centering
\includegraphics[scale=1]{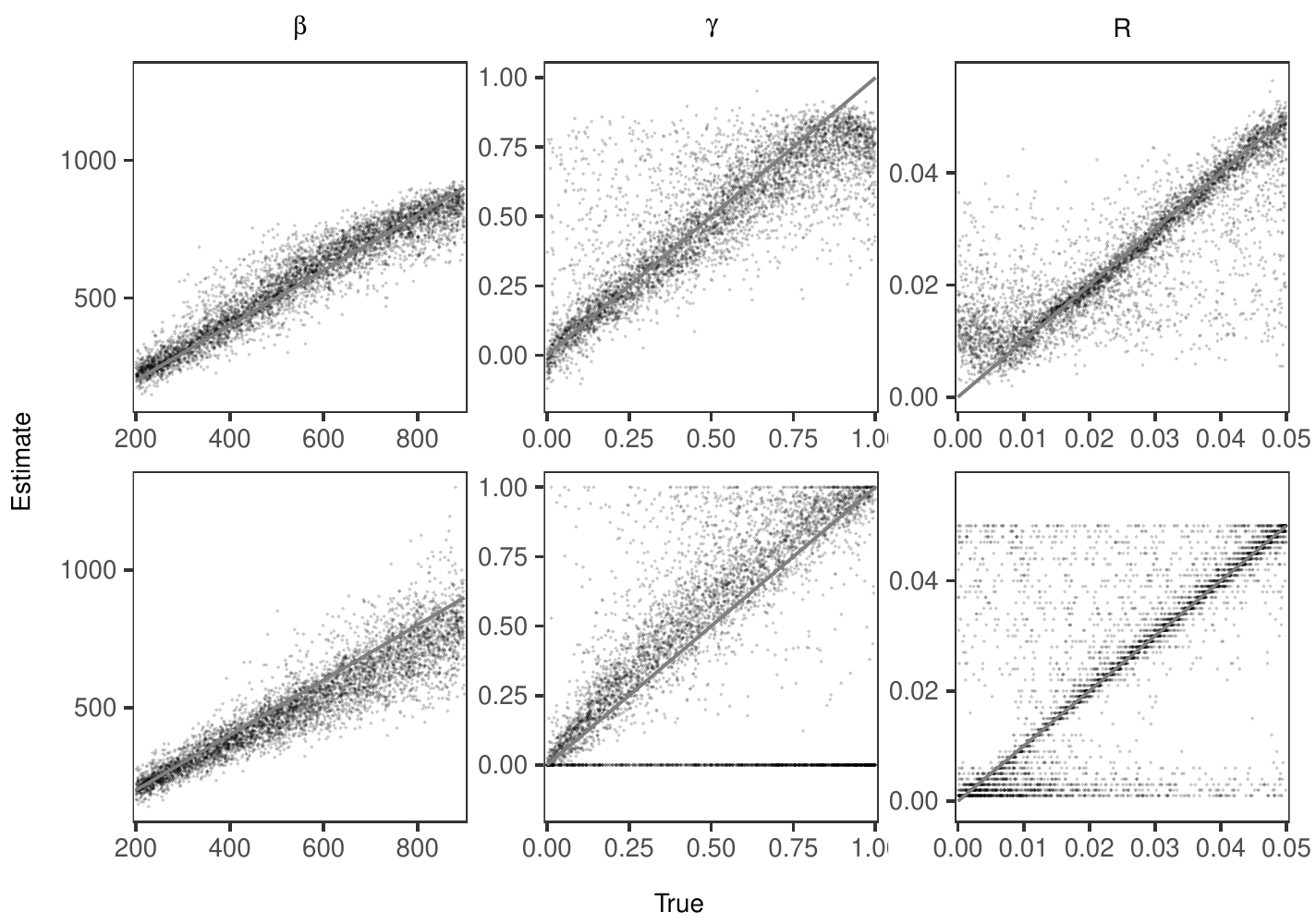}
\caption{Estimated parameters obtained with the neural network approach (top row) and profile maximum pseudo likelihood (bottom row) plotted against true parameters in a Strauss process. The parameter is stated at the top of each column. The solid gray line is the identity line.}
\label{fig:Strauss_estimates}
\end{figure}

\begin{figure}[ht!]
\centering
\includegraphics[scale=1]{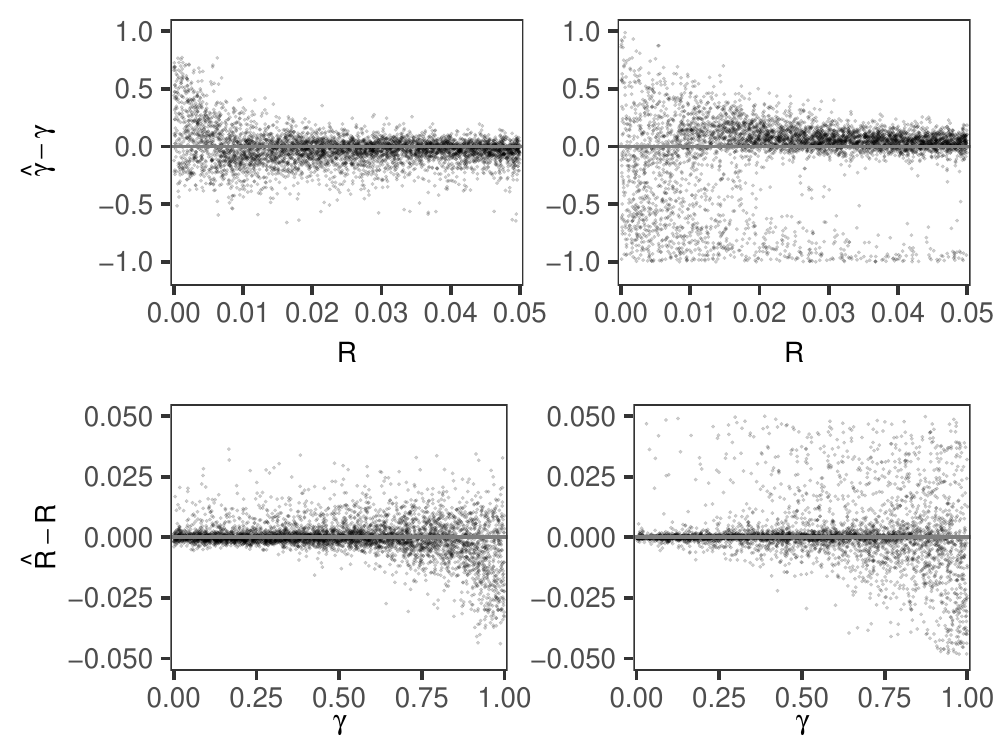}\caption{Estimates of $\gamma$ and $R$ minus their true value plotted against $R$ and $\gamma$, respectively, for a Strauss process. Estimates were obtained with the neural network approach (left) and maximum profile pseudo likelihood estimation (right).}
\label{fig:Strauss_special}
\end{figure}

\begin{figure}[ht!]
\centering
\includegraphics[scale=1]{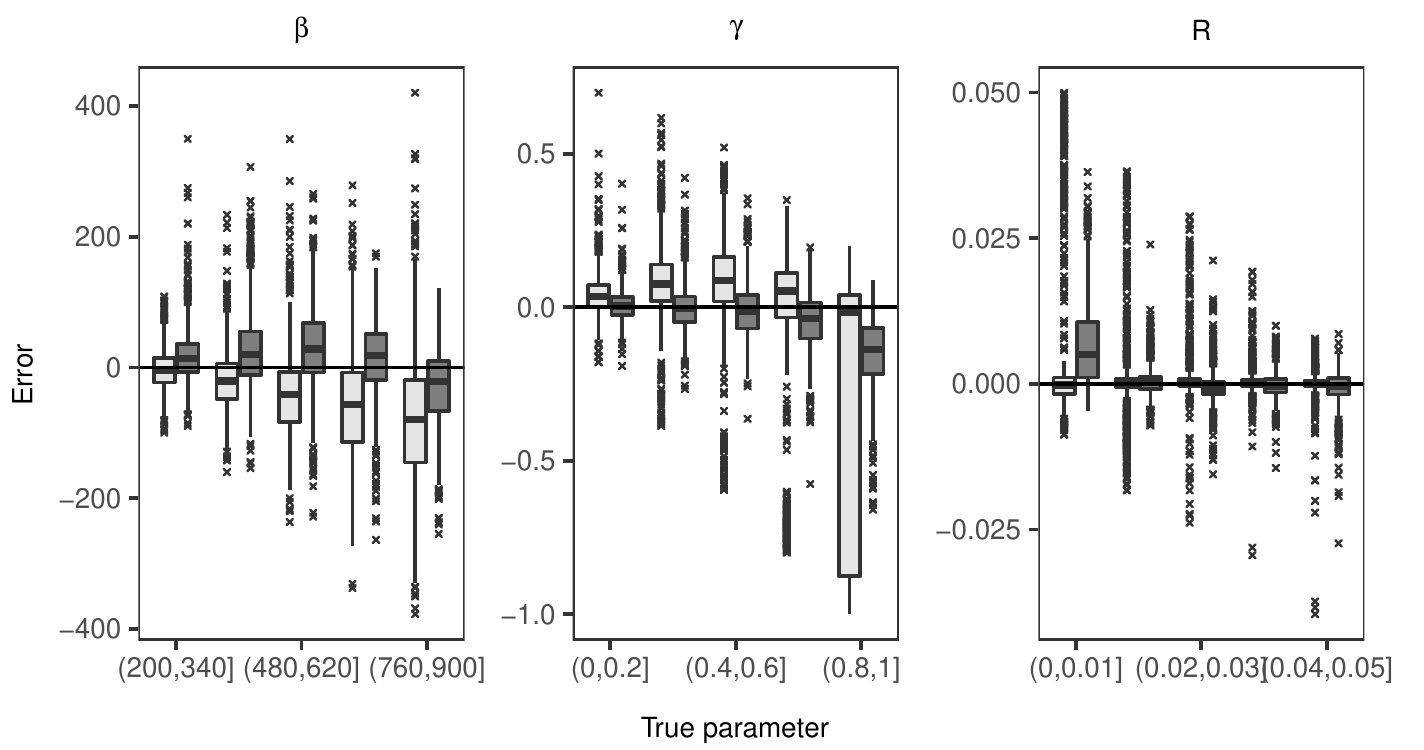}
\caption{Boxplot of the errors (estimated minus true value) for the parameter of a Strauss process stated at the top of the plots. The estimates were obtained with maximum profile pseudo likelihood estimation (light gray) and the neural network approach (dark gray). For the parameters $\gamma$ and $R$, cases where $R<0.01$ and $\gamma > 0.7$, respectively, were omitted.}
\label{fig:Strauss_boxplot}
\end{figure}

For the neural network approach, I used $5{,}000$ simulations of a Strauss process with parameters sampled in the intervals $\beta \in (200, 900)$, $\gamma\in (0,1)$, and $R\in (0, 0.05)$ for the training data and made further $5{,}000$ simulations for a test data set. For justification of the number of simulations in the training data see Appendix~\ref{app:trainingdata}. The simulation of Strauss processes which I used involves Markov chains, and a shared burnin for all simulations was chosen based on trace plots of the number of points and $R$-close pairs for certain combinations of the parameters believed to require the most iterations. Based on this, I used $100{,}000$ iterations of the Markov chain. Figures~\ref{fig:Strauss_estimates}--\ref{fig:Strauss_boxplot} show some plots for the estimated parameters for the point patterns in the test set.
The estimates were obtained with either the neural network approach or profile maximum pseudo likelihood estimation. A clear advantage of the neural network approach is that all parameters can be estimated simultaneously, and the estimate of $R$ is thus not restricted to a finite set of values.

For $\beta$, there is overall less variation in the error obtained with the neural network approach compared to the method of profile maximum pseudo likelihood estimation where the variation increases with the true value of $\beta$, something which does not happen with the neural network approach. The method of profile maximum pseudo likelihood estimation have a tendency to underestimate $\beta$, which gets worse as the true value increases whereas the neural network approach has an overall tendency to slightly overestimate it.

For $\gamma$,  both methods struggle when $R$ is small as seen by Figure~\ref{fig:Strauss_special}, but the neural network approach seems to overall handle it better than profile maximum pseudo likelihood estimation since it can apparently handle the estimation of $\gamma$ well for smaller values of $R$ than profile maximum pseudo likelihood. For a better comparison of the methods, cases where $R < 0.01$ are excluded from the plot for $\gamma$ in Figure~\ref{fig:Strauss_boxplot}. The neural network approach recovers $\gamma$ very well if the true value is not above circa $0.8$, and in this case it also performs better than profile maximum pseudo likelihood estimation which has a tendency for overestimating $\gamma$. If the true value of $\gamma$ is above circa $0.8$, the neural network approach in general underestimates $\gamma$ whereas profile maximum pseudo likelihood estimation either estimates it to be near $1$, as it should, or near $0$. 

For $R$, both methods struggle when $\gamma$ is high, so cases where $\gamma > 0.7$ are excluded from the plot for $R$ in Figure~\ref{fig:Strauss_boxplot}. The neural network approach has difficulties recovering $R$ when the true value is small in which case profile maximum pseudo likelihood estimation shows better performance; however, if the true value is above circa $0.01$, the neural network approach recovers $R$ very well and the results are comparable to those obtained with profile maximum pseudo likelihood estimation. Remember though, that the performance of profile maximum pseudo likelihood estimation depends much on how fine a grid of $R$-values one considers.

\begin{figure}[ht!]
\centering
\includegraphics[scale=1]{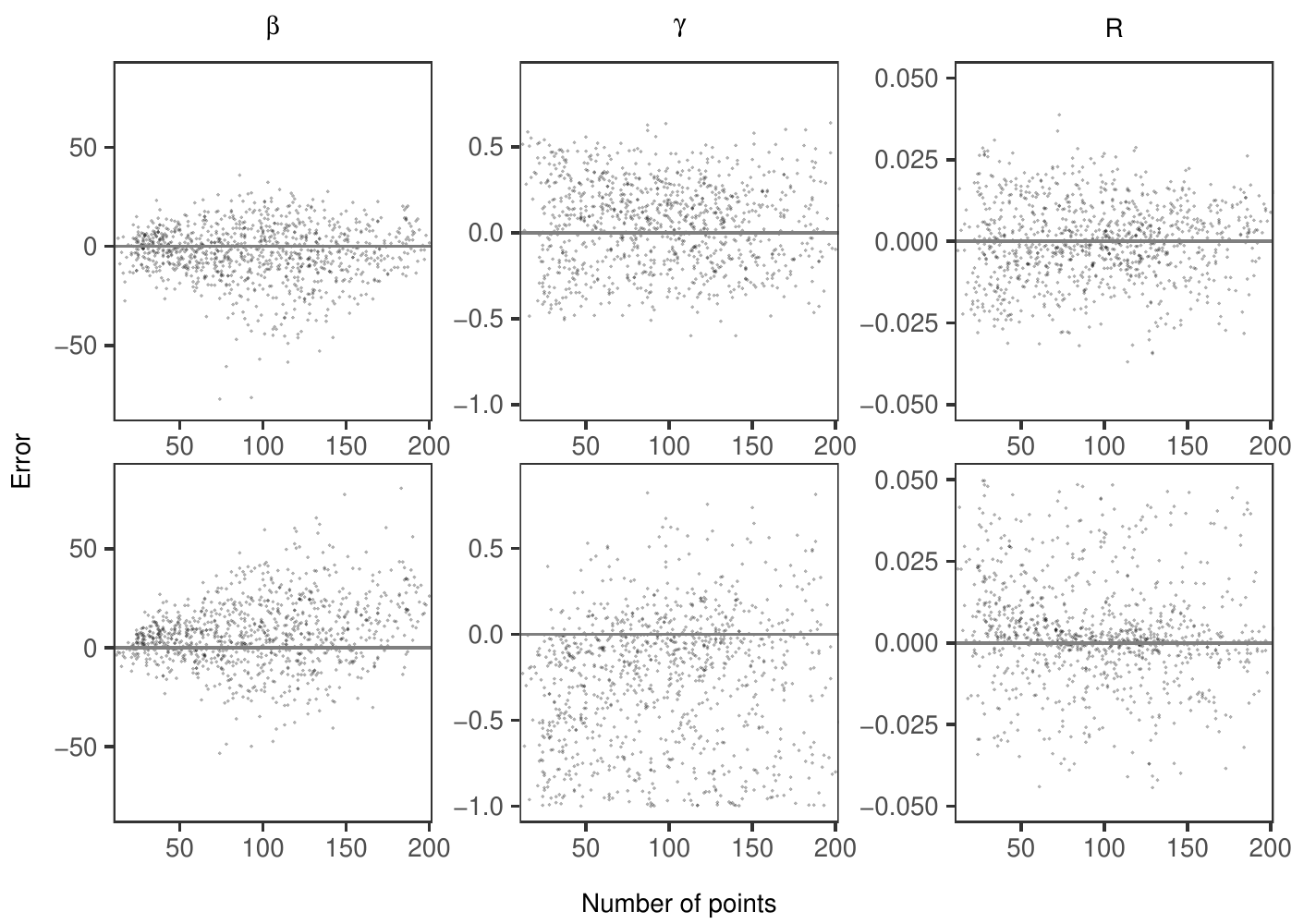}
\caption{Estimated minus true value plotted against the number of points in the point pattern. The parameter of the Strauss process is stated at the top of each column. In the top row, estimates were obtained with the neural network approach; in the bottom row, estimates were obtained with profile maximum pseudo likelihood estimation.}
\label{fig:error_vs_npoints_Strauss}
\end{figure}

In order to asses how the method performs on point patterns with few points, I made a second simulation study where I considered $\beta\in(20,200)$ for the training and test data. I used $1{,}000$ simulations in the test set, and in this case I only trained the network for 10 epochs because a plot like in Figure~\ref{fig:epoch} revealed problems with overfitting when training the network for longer. Everything else was as above. Figure~\ref{fig:error_vs_npoints_Strauss} shows the errors of the estimates obtained with the neural network approach and profile maximum pseudo likelihood estimation for each parameter for the test cases where the number of points was below 200. 
The plots reveal no clear tendencies in the estimates obtained with the neural network approach. In the case of maximum profile pseudo likelihood estimation there are no clear tendencies for $\beta$, but for $\gamma$ there is a general tendency to underestimate and for $R$ there is a tendency to overestimate if there are very few points in the point pattern. 

\subsection{LGCP-Strauss processes}
\label{sec:LGCPStrauss}

An LGCP-Strauss process is a model for repulsion at small scale and clustering at a larger scale. It is a combination of an LGCP and a Strauss process, and, defined on $W$, it has density 
\begin{equation*}
f(\textbf{x})= \mathrm{E}\left[\frac{1}{C_{\theta}(Y)} \exp\left (\sum_{i=1}^n Y(x_i)\right )\gamma^{S_{R}(x)}\right]
\end{equation*}
for $x=\{x_1,\ldots,x_n\}\subset W$ with respect to the unit rate Poisson process where $\theta$ is the parameter vector; $Y=\{Y(u)\}_{u\in W}$ is a Gaussian random field; the expectation is with respect to $Y$; and $C_{\theta}(Y)$ is the normalising constant obtained when conditioning on $Y$. For $Y$, I used a parametrization as in Section~\ref{sec:LGCP} with parameters $\mu$, $\sigma^2$, and $s$, so $\theta = (\mu, \sigma^2, s, \gamma, R)$ where $\gamma \in [0,1]$. If $\gamma = 1$ or $R=0$, it collapses to an LGCP; if $\sigma^2=0$, it collapses to a Strauss process. 

I made the training data for the neural network based on $40{,}000$ simulations of an LGCP-Strauss process with parameters sampled in the intervals $\mu\in (4.5,6)$, $\sigma^2\in (0,4)$, $s\in (0.001,0.1)$, $\gamma \in (0,1)$, and $R\in (0,0.05)$. For justification of the number of simulations in the training data see Appendix~\ref{app:trainingdata}. I made further $5{,}000$ simulations for a test set for which I estimated the parameters with the neural network approach. The simulation of LGCP-Strauss processes which I used involves Markov chains, and a shared burnin for all simulations was chosen based on trace plots of the number of points and $R$-close pairs for certain combinations of the parameters believed to require the most iterations. Based on this, I used $200{,}000$ iterations of the Markov chain.

Regarding estimating the parameters of an LGCP-Strauss process \citet{LGCPStrauss} noted that the usual methods for estimating parameters in point process models are intractable for this model and thus used ABC. I therefore compare the estimates obtained with the neural network approach to approximate posterior means obtained with an ABC technique. Specifically, I used the method of ABC via random forests as implemented in the \texttt{R}-package \texttt{abcrf} version 1.8.1 \citep{abcrf}. In short, this method trains a regression random forest on a reference table consisting of chosen summary statistics calculated for a number of prior predictions with the aim of predicting posterior expectations, variances and quantiles for a parameter. A regression random forest consists of a number of regression trees trained on bootstrap samples of the training data. In each regression tree the input is subjected to a number of binary decision rules after which a leaf of the tree will be reached. The prediction made by this regression tree is then the mean of the response variables from its training data which are associated to this leaf. The prediction of the random forest is then the mean of the predictions from each individual tree. I refer to \citet{abcrf_paper} for more details about ABC via random forests.

In this paper, I am only interested in the approximate posterior means obtained with ABC via random forests, and these are the predictions of the parameters made by trained random forests. The approach is thus quite similar to the neural network approach except that the training data is used to train random forests instead of a neural network. As recommended in \citet{abcrf_paper}, I trained an independent random forest for each parameter, and the parameters are thus estimated separately instead of simultaneously as with the neural network approach. I used random forests with 500 trees (the default in \texttt{abcrf}) and made the check recommended in \citet{abcrf_paper} for whether this was sufficient.  As a reference table, I used the same training data as for the neural network since both methods are based on machine learning techniques and I wanted to compare their performance when given the exact same information; for the same reason I did not investigate whether the ABC technique would benefit from more simulations in the training data. So in the ABC approach, the independent uniform distributions used to sample the parameters for the training data serve as prior distributions.

\begin{figure}[ht!]
\centering
\includegraphics[scale=1]{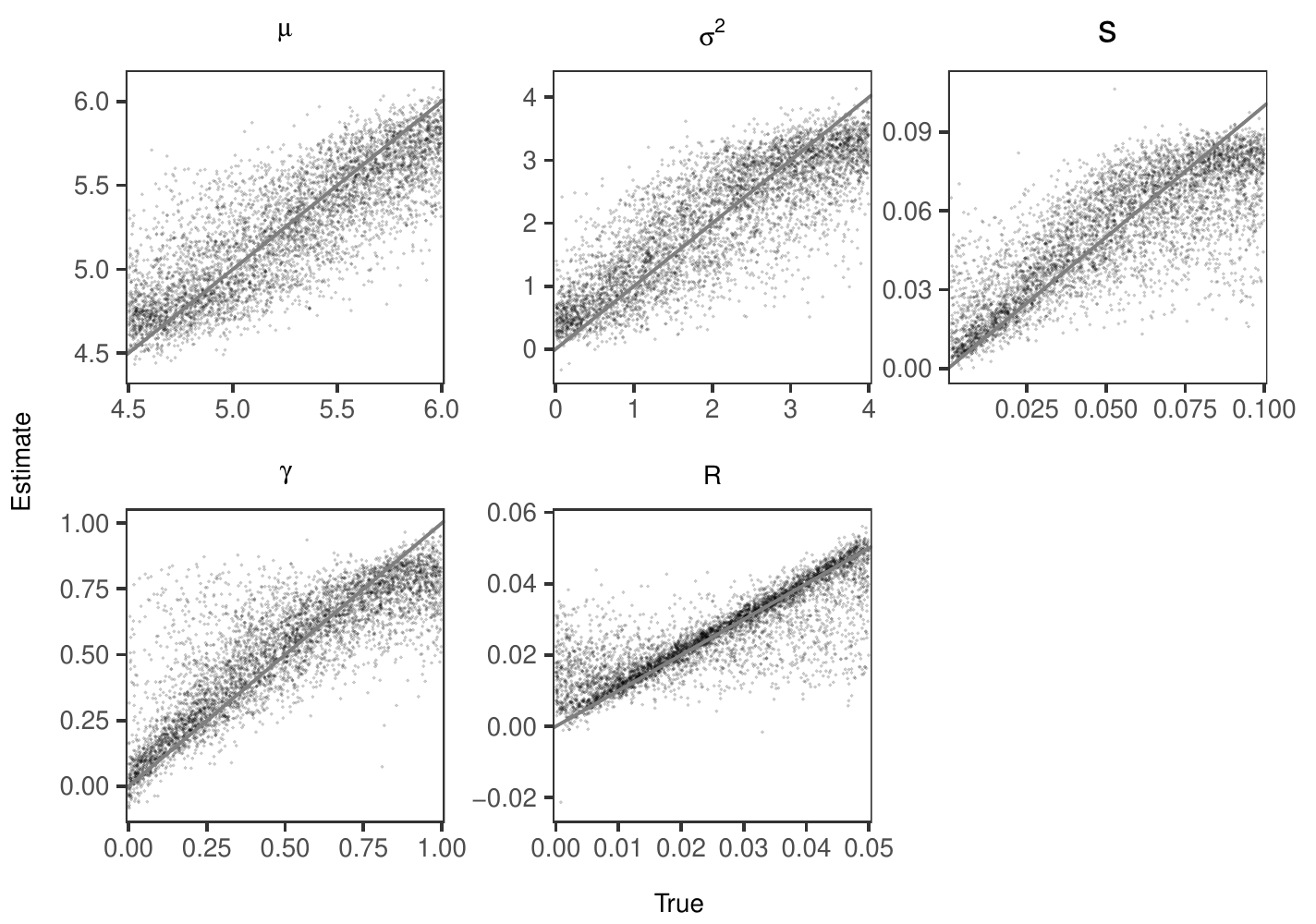}
\caption{Estimated parameters obtained with the neural network approach plotted against true parameters in an LGCP-Strauss process. The parameter is stated at the top of each plot. The solid gray line is the identity line.}
\label{fig:LGCPStrauss_estimates_NNet}
\end{figure}

\begin{figure}[ht!]
\centering
\includegraphics[scale=1]{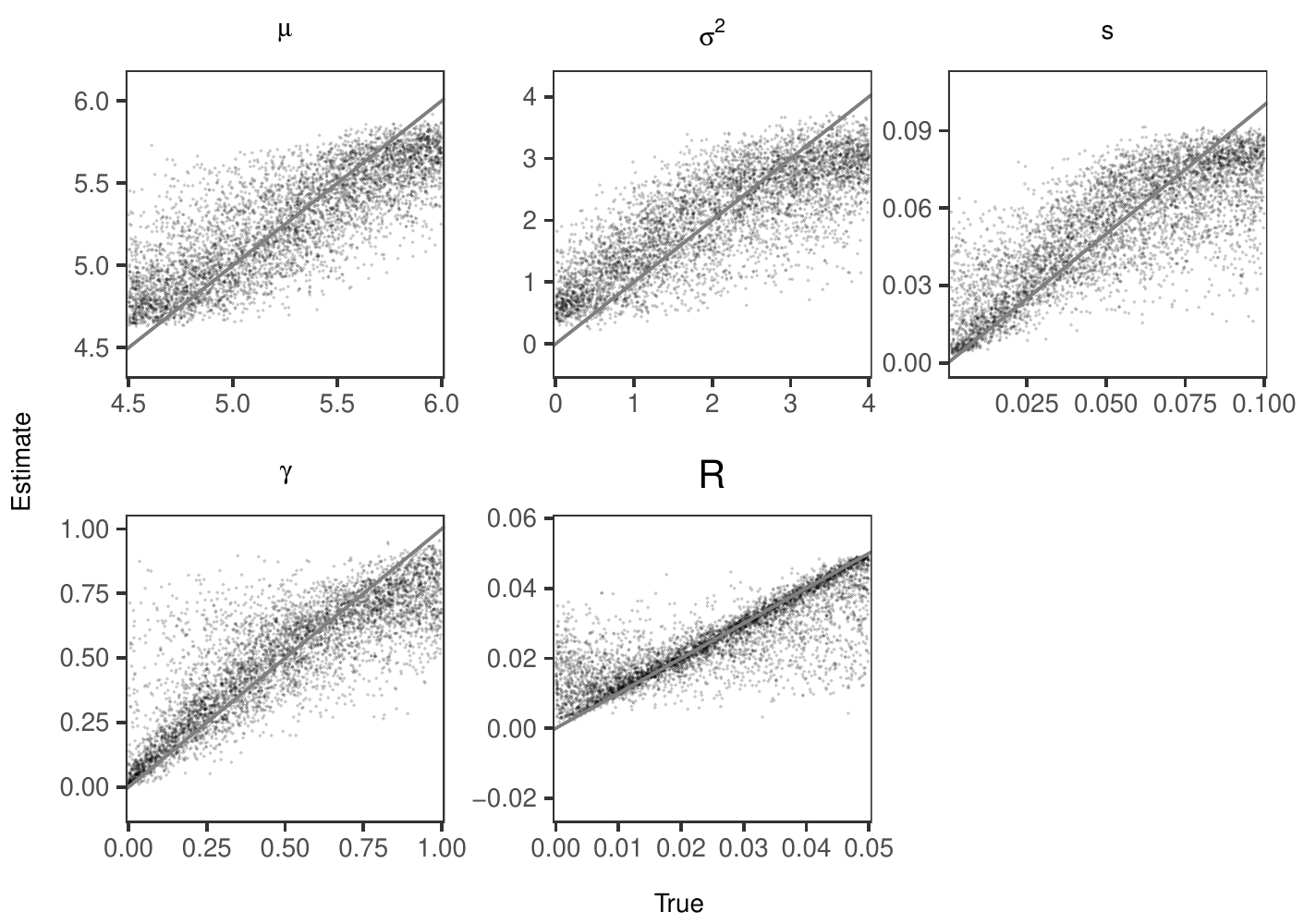}
\caption{Approximate posterior means obtained with ABC via random forests plotted against true parameters in an LGCP-Strauss process. The parameter is stated at the top of each plot. The solid gray line is the identity line.}
\label{fig:LGCPStrauss_estimates_ABC}
\end{figure}

\begin{figure}[ht!]
\centering
\includegraphics[scale=1]{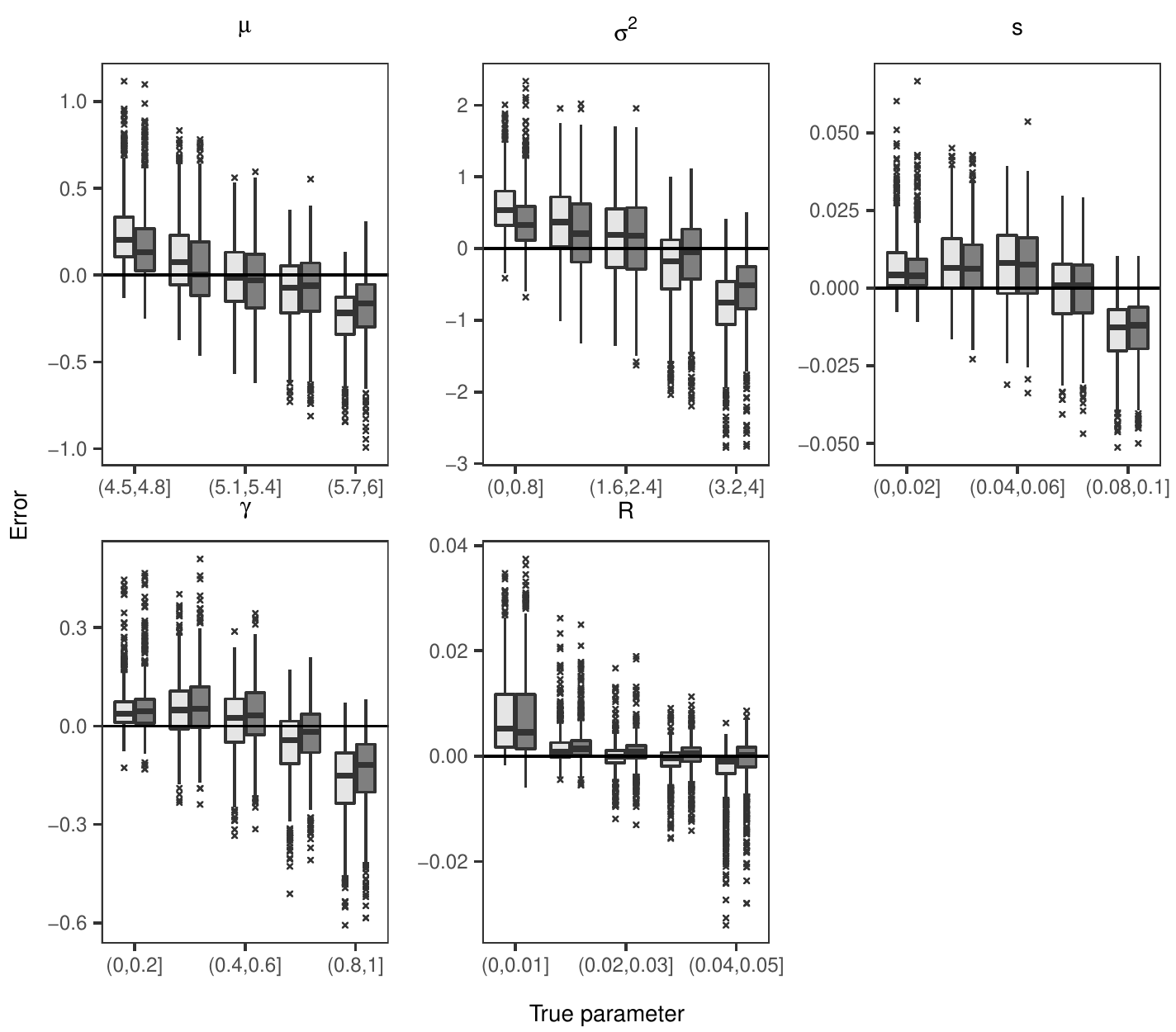}
\caption{Boxplot of the errors (estimated minus true value) for the parameter of an LGCP-Strauss process stated at the top of each plot. The estimates were obtained with an ABC technique (light gray) and the neural network approach (dark gray), respectively. For the parameters $s$, $\gamma$, and $R$, cases where $\sigma^2>1$, $R<0.01$ and $\gamma > 0.7$, respectively, were omitted.}
\label{fig:LGCPStrauss_boxplot}
\end{figure}

The estimates obtained with the neural network approach and the posterior means obtained with ABC via random forests are plotted against the true parameter values in Figures~\ref{fig:LGCPStrauss_estimates_NNet}--\ref{fig:LGCPStrauss_estimates_ABC}.
Figure~\ref{fig:LGCPStrauss_boxplot} shows boxplots of the errors where some cases are omitted in the plots for $s$, $\gamma$, and $R$ due to arguments similar to those in Sections~\ref{sec:LGCP} and \ref{sec:strauss}.
The results obtained with the two methods are very similar, except that the neural network approach performs slightly better for $\mu$ and $\sigma^2$ near the endpoints of the considered intervals. It is apparently easiest to estimate $\gamma$ and $R$, which are recovered very well except that there is again a tendency to underestimate $\gamma$ when the true value is high and to overestimate $R$ when the true value is small. The estimate for $s$ is again best when the true value is small. There is a tendency for overestimating $\sigma^2$ unless the true value is above circa $3$ in which case it is usually underestimated. \citet{LGCPStrauss} also found it to be difficult to make inference about the parameters of the Gaussian random field in an LGCP-Strauss process and related it to the fact that it can be difficult to see the effect of changes in the Gaussian random field from a realization of the process because it is obscured by the small scale regularity.

The LGCP-Strauss process models quite complex behaviour in point patterns, so I do not think it is appropriate to fit it to point patterns with few points. I therefore do not consider a second simulation study focusing on point patterns with few points as I did in Sections~\ref{sec:LGCP}--\ref{sec:strauss}.

\subsection{Some remarks about speed}
The purpose of this section is to give an idea of how time consuming the neural network approach is even though this will of course depend heavily on implementation, software, the data a model should be fitted to etc. All the below timings were made using just a single core, but some of the calculations can also be run in parallel. 

In the situations in Sections~\ref{sec:LGCP}--\ref{sec:LGCPStrauss}, it took about $5.4,17$, and $115$ minutes, respectively, to make the training data and $2.6,1.5$, and $9.7$ minutes, respectively, to train the neural network. The most time consuming part of the procedure is to make the training data, especially to make the simulations. However, the process of making simulations and calculating summary statistics can easily be parallelized if multiple cores are available. I also recommend to always take the extra time to make a test data set which can be used to asses the performance of the method in a given situation. After the network was trained, it took about 1 second in all three cases to fit the point process model to the $5{,}000$ simulations in the test data. 

With the method of minimum contrast estimation it took about $100.8$ minutes to fit LGCP models to the $5{,}000$ simulations in the test data in Section~\ref{sec:LGCP}, and it thus took $1.21$ seconds on average to fit one model; with the method of profile maximum pseudo likelihood estimation it took about 154 minutes to fit Strauss process models to the $5{,}000$ simulations in the test data in Section~\ref{sec:strauss}, and it thus took $1.85$ seconds on average to fit one model (these timings of course depend heavily on how many values of the parameter $R$ one considers). When fitting a single model, it is thus much faster to use minimum contrast estimation or profile maximum pseudo likelihood estimation than to use the neural network approach. However, after the neural network has been trained, it can be used to fit the spatial point process model to multiple point patterns as long as they are well represented in the training data, and this can be done very fast. If a model is to be fitted to multiple point patterns and it is possible to train a neural network which is suitable for all these cases, the neural network approach can be faster. With the ABC method in Section~\ref{sec:LGCPStrauss} it took about $93.4$ minutes to fit the random forest objects and $34$ minutes to make the predictions which include predictions of the posterior means. The ABC procedure also needs the time for making the training data. Thus, the neural network approach was faster in this case, but it is possible to use parallelization in the ABC method in order to speed it up. 

\section{Data example}\label{sec:dataex}

\begin{figure}[ht!]
\centering
\includegraphics[scale=1]{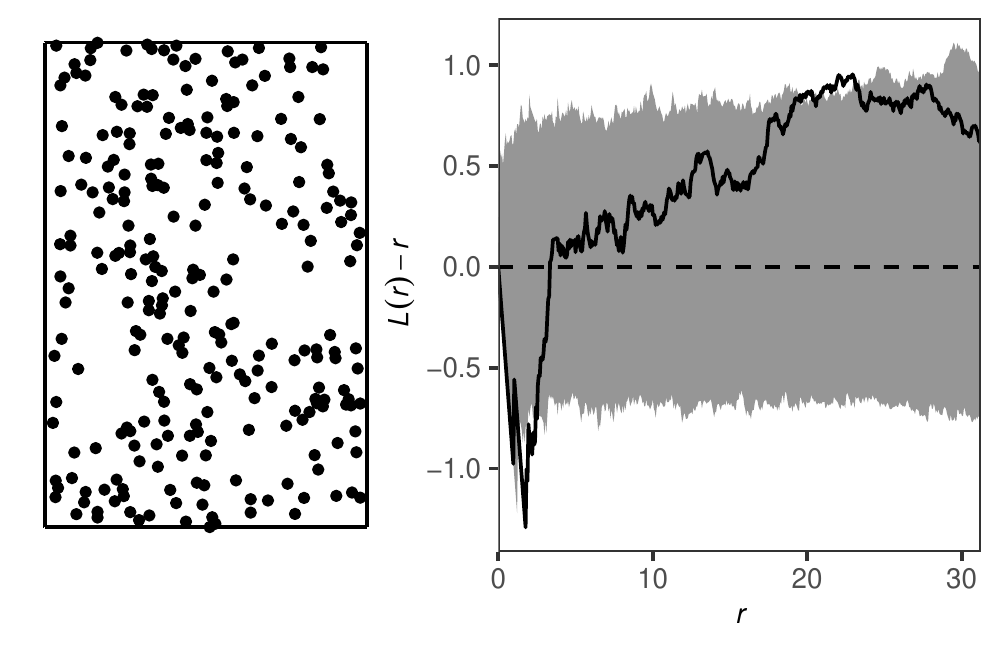}
\caption{Left: Point pattern of the locations of 256 oak trees which suffer from frost shake in a $125\times 188$ m rectangular region of Allogny in France. Right: $\hat{L}(r)-r$ together with a $95\%$ global envelope calculated from 2499 simulations of a homogeneous Poisson process.}
\label{fig:data}
\end{figure}

The left panel in Figure~\ref{fig:data} shows the part of the Allogony data set from the R-package \texttt{ads} version 1.5-5 \citep{ads} which contains the locations of 256 oak trees which suffer from frost shake in a $125\times 188$ m rectangular region of Allogny in France (this rectangular region is $W$). The right panel shows $\hat{L}(r)-r$ together with a $95\%$ global envelope for the null hypothesis that data comes from a homogeneous Poisson process. 
Briefly, a 95\% global envelope is a region for which the functional summary statistic calculated from the observed data will fall completely within if and only if the null hypothesis cannot be rejected at level approximately $5\%$. The envelope was calculated from 2499 simulations (thereby following the recommended number of simulations in \citet{GET2017}) of a homogeneous Poisson process and based on the extreme rank length (see \citet{GET2017, GET2018, GETinR} for more information about global envelopes and the \texttt{R}-package \texttt{GET} (version 0.2-4), which I used to calculate them). The plots indicate that the point pattern exhibit repulsive behaviour at a small scale and some clustering at a larger scale. As an example, I now show how the neural network approach can be used to fit an LGCP-Strauss process to this point pattern (an LGCP-Strauss process model was previously fitted to this data in \citet{LGCPStrauss}).

\begin{figure}[ht!]
\centering
\includegraphics[scale=1]{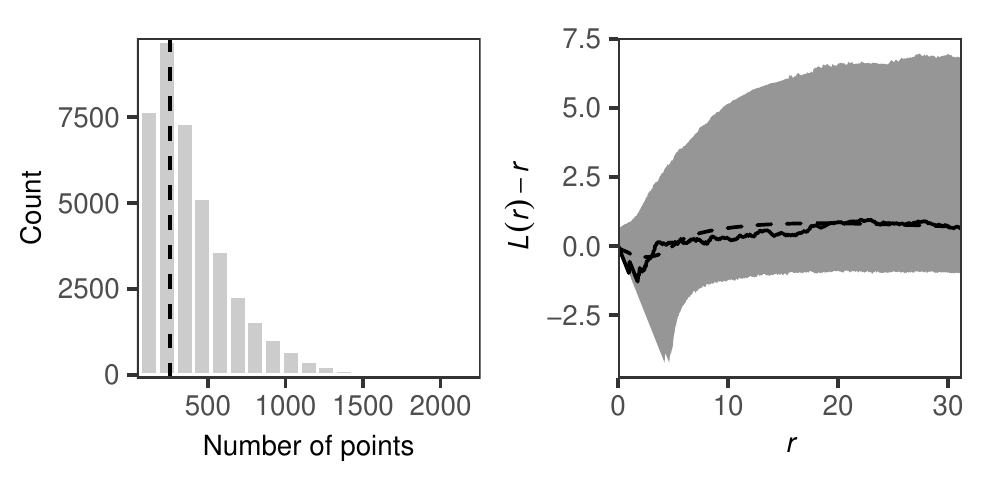}
\caption{Left: Histogram of the number of points in the simulations in the training data where the vertical dashed line indicates the number of points in the observed point pattern of oak trees. Right: A 95\% global envelope calculated from the $40{,}000$ estimates of $L(r)-r$ in the training data (gray area), the mean (dashed curve) and $\hat{L}(r)-r$ obtained from the observed point pattern of oak trees (solid curve).}
\label{fig:n_L_train}
\end{figure}

Regarding the ranges of parameters to use in the training data for the neural network, I used $\gamma\in(0,0.7)$ and $R \in (1,5)$ since Figure~\ref{fig:data} shows clear evidence of repulsion in the observed point pattern and \citet{LGCPStrauss} noted that the interaction radius $R$ is often near the $r$-value which gives the smallest value of $\hat{L}(r)-r$. For the parameters of the Gaussian random field, I decided to use $\mu\in (-5.6,-3)$, $\sigma^2\in(0,2)$, and $s\in (0.001, 15)$ after having looked at some simulations of LGCP-Strauss processes. I then used $40{,}000$ simulations on $W$ where the parameters were sampled uniformly on the above intervals to construct the training data for the neural network approach. Figure~\ref{fig:n_L_train} shows a histogram of the number of points in the point patterns in the training data and a 95\% global envelope calculated from the $40{,}000$ estimates of $L(r)-r$ in the training data. The same summaries obtained from the oak point pattern are also shown in the plots, where we see that both the observed number of points and the behaviour of $\hat{L}(r)-r$ are well represented in the training data, which is crucial in order to get reliable estimates with the neural network approach. A check like this may both reveal if the intervals for the parameters have been chosen inappropriately or if the considered class of model is ill-suited for fitting the observed point pattern.

\begin{figure}[ht!]
\centering
\includegraphics[scale=1]{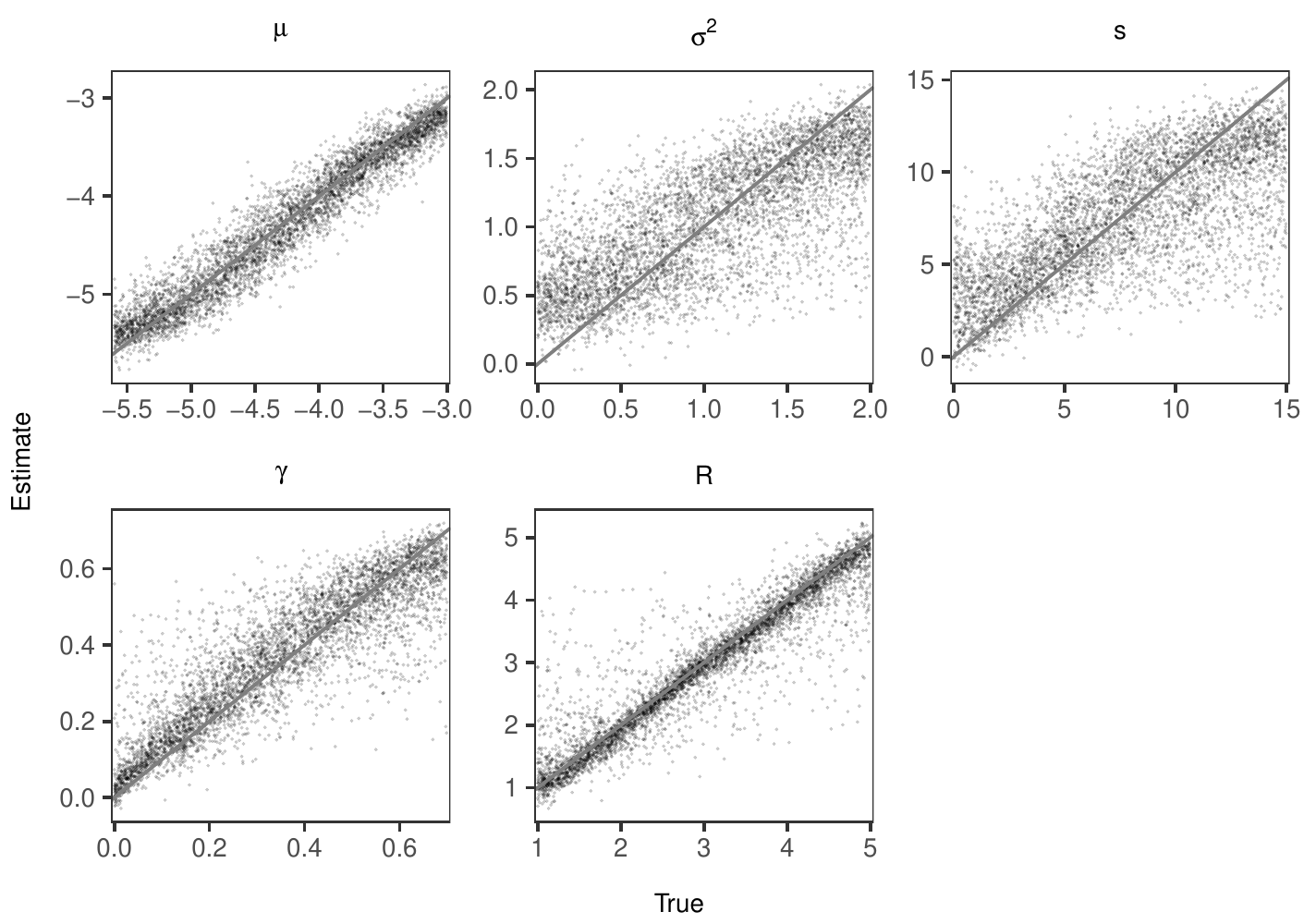}
\caption{Estimated parameters of the test data obtained with the neural network approach aimed at fitting an LGCP-Strauss process to the oak point pattern plotted against true parameters. The parameter is stated at the top of each plot.}
\label{fig:estimates_dataexample}
\end{figure}

I also made $5{,}000$ simulations for a test data set, and Figure~\ref{fig:estimates_dataexample} shows the estimated parameters for these plotted against the true values. This shows that in this situation $\mu, \gamma,$ and $R$ are recovered well whereas there is more uncertainty in the estimates of $\sigma^2$ and $s$.

When using the trained neural network to estimate the parameters for the point pattern of oak trees, I got the estimates $\hat{\mu} = -4.54, \hat{\sigma}^2 = 0.32, \hat{s} = 10.93, \hat{\gamma} = 0.21,$ and $\hat{R} = 1.91$. The most popular way to validate a fitted point process model is to consider global envelopes and corresponding tests calculated for some functional summary statistic. I did not want to use the $L$-function for this global envelope and test since it plays a major part in the estimating procedure. I therefore used the $J$-function given in \eqref{eq:J} instead. I used the non-parametric estimate $ \widehat{J}(r)=(1-\widehat{G})/(1-\widehat{F})$ where $\widehat{G}$ and $\widehat{F}$ are the so-called Kaplan-Meier estimators of $G$ and $F$, which account for edge effects, see \citet[Section 8.11.4]{spatstat} for how these estimators are given. Regarding the considered range of $r$-values for $\widehat{J}(r)$, the function \texttt{Jest} from \texttt{spatstat} which is used to estimate $J$ gives a recommendation, which I have followed.

\begin{figure}[ht!]
\centering
\includegraphics[scale=1]{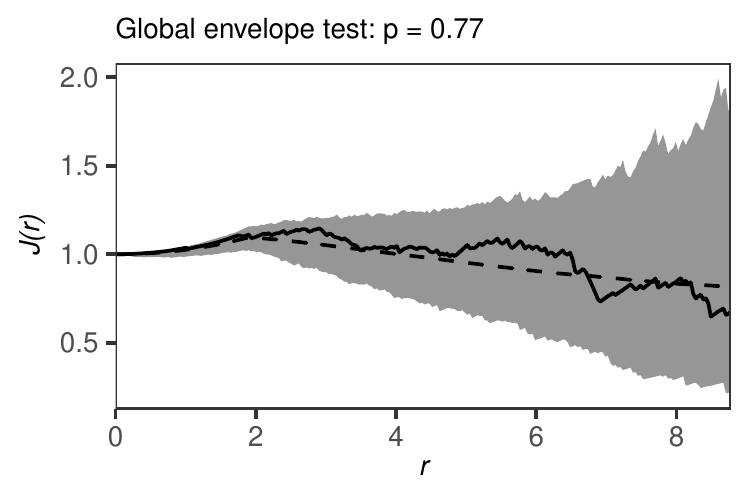}
\caption{A 95\% global envelope based on the $J$-function calculated from $2499$ simulations under the LGCP-Strauss process model fitted to the oak point pattern  (gray area), the mean obtained from the simulations (dashed curve), and the estimate calculated from the observed point pattern (solid curve). The $p$-value of the corresponding global envelope test is stated at the top.}
\label{fig:envelope_fit}
\end{figure}

Figure~\ref{fig:envelope_fit} shows a 95\% global envelope and the $p$-value of the corresponding global envelope test based on the $J$-function and calculated from $2499$ simulations under the fitted model. This indicates that the fitted model describes the point pattern of oak trees very well. 

\section{Discussion and future research}
I have presented a method which is generally applicable to estimate  parameters in all spatial point process models which it is possible to simulate from. The method recovers parameters well compared to common estimating techniques since it gives either better or similar results. The advantages of the method are that the only necessary information about the model is a tractable simulation procedure and that all unknown parameters can be estimated simultaneously. The method is more time consuming than minimum contrast estimation and profile maximum pseudo likelihood estimation when it comes to fit a single model. However, the most time consuming part of the method is to make training data and to train the neural network, so if it is possible to train a neural network which can be reused to fit a model to many point patterns, the neural network approach can be faster than using minimum contrast estimation or maximum pseudo likelihood estimation on each point pattern. Compared to ABC, the neural network approach is also potentially faster. 

Future research may include how to use the neural network approach to estimate parameters in inhomogeneous point process models which include covariate information. It could also be interesting to explore the possibility to pre-train large neural networks which could be applicable to a wide range of point patterns which are often encountered in practice thereby obtaining a very fast estimation procedure for such point patterns.  

\section*{Acknowledgements} The research of the author was supported by The Danish Council for Independent Research | Natural Sciences, grant DFF -- 7014-00074 `Statistics for point processes in space and beyond'. I would also like to thank Jesper Møller for his helpful comments regarding this paper.

\appendix

\section{Details for the neural network approach}
\label{app:NNet_details}
\subsection{Training data}\label{app:trainingdata} 

\begin{figure}[ht!]
\centering
\includegraphics[scale=1]{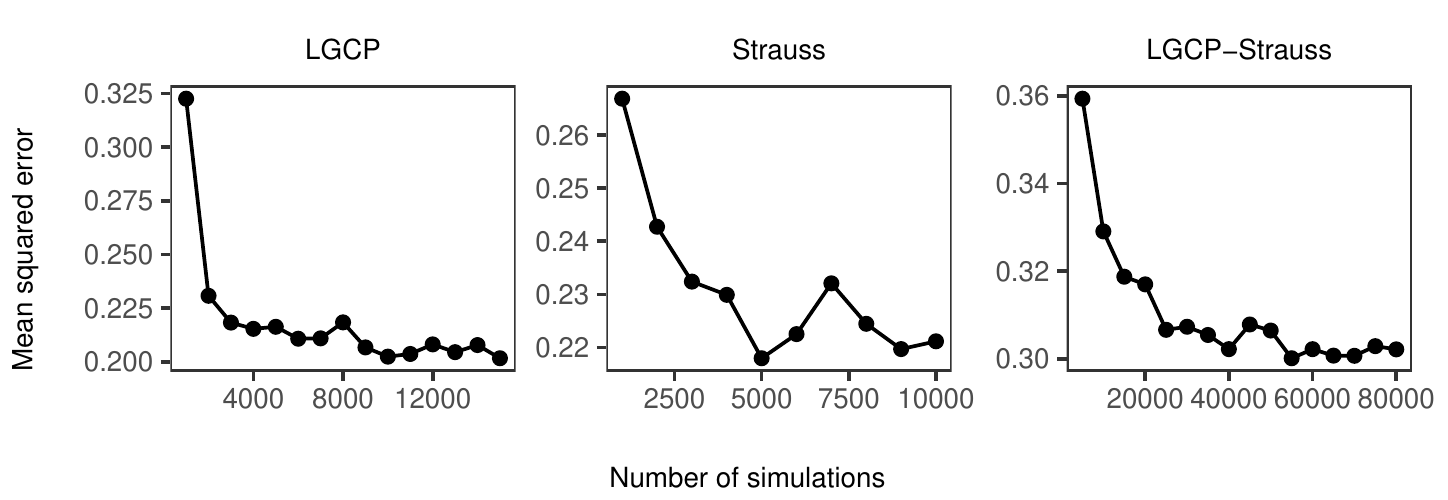}
\caption{Mean squared errors calculated for test sets with $5{,}000$ simulations for the situations of the first simulation studies in Sections~\ref{sec:LGCP}--\ref{sec:LGCPStrauss} as stated at the top plotted against the number of simulations in the training data.}
\label{fig:nsim}
\end{figure}

It is possible to get as much training data as desired since it is merely a matter of making more simulations, but simulation procedures may be time consuming. Figure~\ref{fig:nsim} shows the mean squared errors obtained with the neural network approach for test sets with $5{,}000$ simulations in the situations of the first simulation studies in Sections~\ref{sec:LGCP}--\ref{sec:LGCPStrauss} plotted against the number of simulations in the training data.
The necessary number of simulations depends on how complicated the model is. Based on Figure~\ref{fig:nsim} I used  $10{,}000$ simulations in the case of an LGCP in Section~\ref{sec:LGCP};  $5{,}000$ simulations in the case of a Strauss process in Section~\ref{sec:strauss}; and $40{,}000$ simulations in the case of an LGCP-Strauss process in Section~\ref{sec:LGCPStrauss}.

\begin{figure}[ht!]
\centering
\includegraphics[scale = 1]{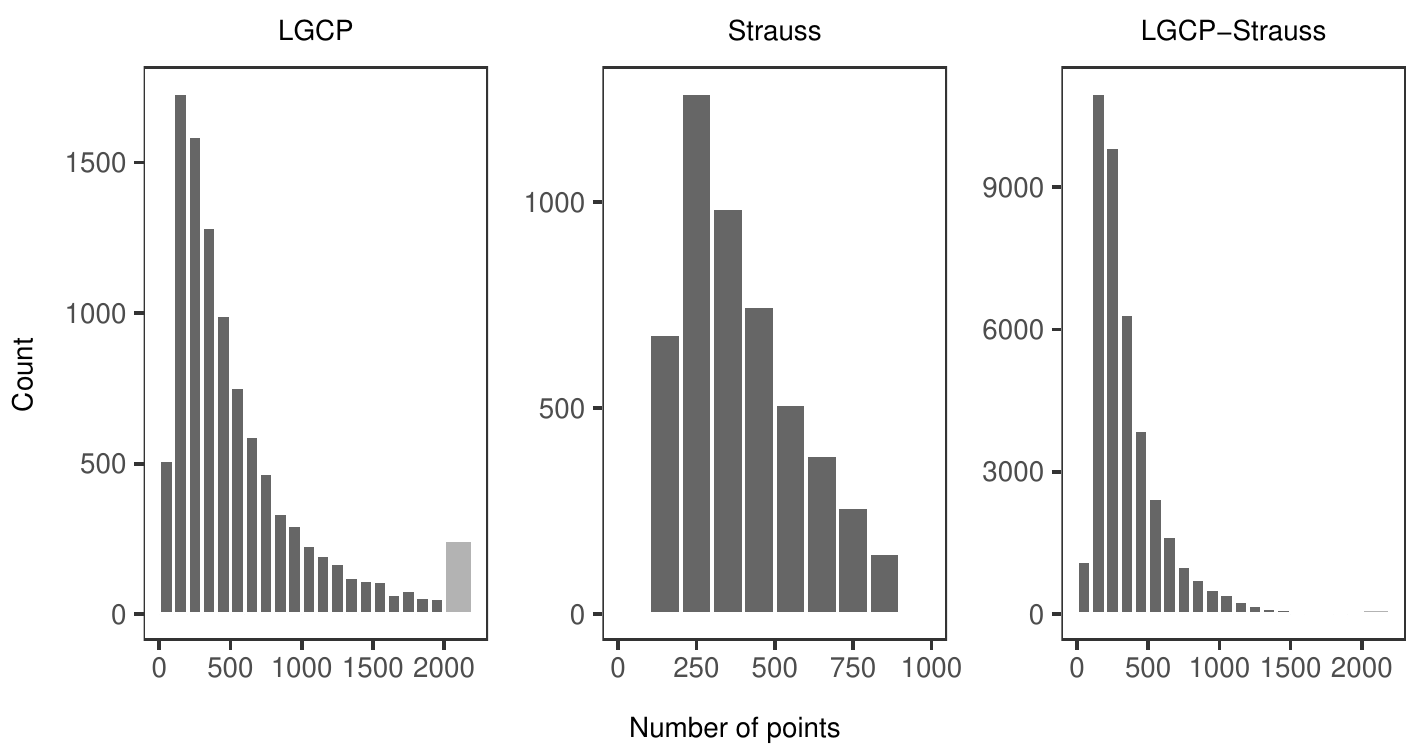}
\caption{Histogram of the number of points in the point patterns in the training data used in the first simulation studies in Sections~\ref{sec:LGCP}--\ref{sec:LGCPStrauss} as stated at the top of each plot. The light grey column indicates the count of point patterns where the number of points is above 2000, the highest value being 6116 and 6497 in the cases of the LGCP and LGCP-Strauss processes, respectively.}
\label{fig:npoints}
\end{figure}

Figure~\ref{fig:npoints} shows histograms of the number of points in the training data sets used for the first simulation studies in Sections~\ref{sec:LGCP}--\ref{sec:LGCPStrauss}.

\subsection{Network training}\label{app:NN_learn}

\begin{figure}[ht!]
\centering
\includegraphics[scale=1]{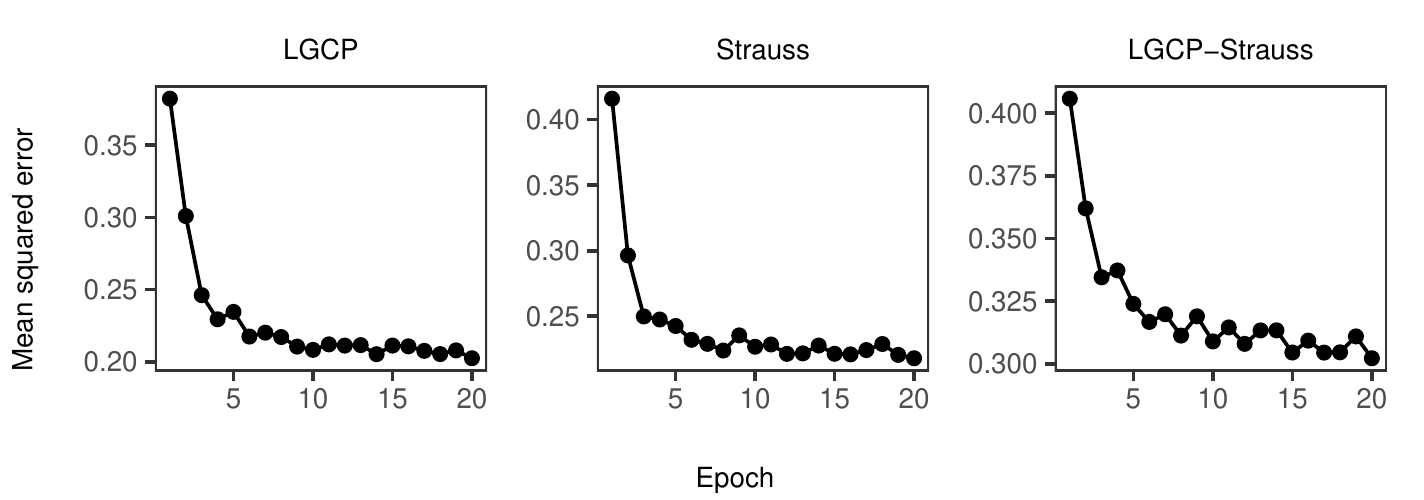}
\caption{Mean squared error calculated from test sets with $5{,}000$ simulations plotted against number of epochs for the situations of the first simulation studies in Sections~\ref{sec:LGCP}--\ref{sec:LGCPStrauss} as stated at the top of each plot. The number of simulations in the training data were as in Sections~\ref{sec:LGCP}--\ref{sec:LGCPStrauss}.}
\label{fig:epoch}
\end{figure}

The unknown parameters of the neural network should be learned based on the training data. This is done by minimizing a loss function with some optimization technique. I used the mean squared error as loss function and the Adam optimizer \citep{adam} for optimization. During training, the training data was send through the network in smaller batches of size 100. An iteration over the entire training data is referred to as an epoch. During training, I also monitored the mean squared error of a test set again constructed from simulations of the point process model as described in Section~\ref{sec:NN_method}. The mean squared error of the test set was among other things used to decide on the number of epochs where the choice in general fell on 20 epochs based on Figure~\ref{fig:epoch}, which also revealed that with the choices I made, there is no apparent problem with overfitting.

\bibliography{bib}

\end{document}